\documentclass[%
 aip,
 amsmath,amssymb,
 reprint,%
]{revtex4-1}

\usepackage{graphicx}
\usepackage{dcolumn}
\usepackage{bm}

\usepackage[utf8]{inputenc}
\usepackage[T1]{fontenc}
\usepackage{mathptmx}
\usepackage{etoolbox}
\usepackage{algorithm}
\usepackage{algorithmicx}
\usepackage[noend]{algpseudocode}
\usepackage{makecell}
\usepackage{xcolor}
\usepackage{caption}                
\usepackage{subcaption}
\usepackage{color}

\makeatletter
\def\@email#1#2{%
 \endgroup
 \patchcmd{\titleblock@produce}
  {\frontmatter@RRAPformat}
  {\frontmatter@RRAPformat{\produce@RRAP{*#1\href{mailto:#2}{#2}}}\frontmatter@RRAPformat}
  {}{}
}%
\makeatother
\begin{document}

\preprint{AIP/123-QED}

\title[Riemann Solver Approach in PINNs]{An approximate Riemann Solver Approach in Physics-Informed Neural Networks for hyperbolic conservation laws}
\author{Jorge F. Urbán}
\affiliation{Departament de Física Aplicada, Universitat d'Alacant, Ap. Correus 99, E-03080 Alacant, Spain}
 \email{jorgefrancisco.urban@ua.es}
\author{José A. Pons}%
\affiliation{Departament de Física Aplicada, Universitat d'Alacant, Ap. Correus 99, E-03080 Alacant, Spain}%

\date{\today}

\begin{abstract}
This study enhances the application of Physics-Informed Neural Networks (PINNs) for modeling discontinuous solutions in both hydrodynamics and relativistic hydrodynamics. 
Conventional PINNs, trained with partial differential equation residuals, frequently face convergence issues and lower accuracy near discontinuities. To address these issues, we build on the recently proposed Locally-Linearized PINNs, which improve shock detection by modifying the Jacobian matrix resulting from the linearization of the equations, only in regions where the velocity field exhibits strong compression. However, the original LLPINN framework required a priori knowledge of shock velocities, limiting its practical utility.
We present a generalized LLPINN method that dynamically computes shock speeds using neighboring states and applies jump conditions 
through entropy constraints. Additionally, we introduce Locally-Roe PINNs, which incorporate an approximate Roe Riemann solver to 
improve shock resolution and conservation properties across discontinuities. These methods are adapted to two-dimensional Riemann problems by using a divergence-based shock detection combined with dimensional splitting, delivering precise solutions.
Compared to a high-order Weighted Essentially Non-Oscillatory (WENO) solver, our method produces sharper shock transitions but smoother solutions in areas with small-scale vortex structures. Future research will aim to improve the resolution of these small-scale features without compromising the model's ability to accurately capture shocks.

\end{abstract}

\maketitle

\section{Introduction}

Solving the governing equations of fluid dynamics is particularly challenging, not only due to the presence of discontinuities such as shock waves and contact surfaces, but also because of the emergence of small-scale vortex structures associated with turbulence \cite{Andreopoulos2000,Durbin2018,Saurel2018}. 
To overcome all these challenges, a wide range of advanced numerical techniques have been developed over the past few decades. 
A group of methods specifically developed to handle discontinuous solutions is categorized as
\emph{high-resolution shock-capturing} (HRSC) methods. Particular well-known examples are the Godunov method \cite{Godunov1959}, the Monotonic Upstream-centered Scheme for Conservation Laws\cite{Vanleer1,Vanleer2} (MUSCL), the Essentially Non-Oscillatory \cite{HARTEN1987231} (ENO) and the Weighted ENO \cite{LIU1994200} (WENO) methods. 

At the heart of these methods lies the Riemann problem, which involves an initial value problem for hyperbolic conservation laws where the initial conditions feature piecewise constant states separated by a discontinuity. This problem is a crucial benchmark, capturing both the fundamental physical behavior and the mathematical structure of the governing conservation laws \cite{toro2009riemann}.
Most numerical schemes use approximate solutions to Riemann problems, solving a Riemann problem at each cell interface to estimate numerical fluxes, which are then used to update the solution in each cell.
 
Additionally, traditional numerical schemes encounter other difficulties. These approaches depend on a spatio-temporal grid (or a basis of orthogonal polynomials for spectral methods)\cite{LeVeque_2002,CanutoSpectral}, which can become very computationally expensive in higher-dimensional simulations due to the increased number of computational cells or spectral coefficients. Handling complex geometries is also difficult, as creating and adapting grids can also be computationally expensive \cite{Mavriplis1997}. Additionally, implementing boundary conditions requires careful attention to avoid numerical artifacts, such as reflections or oscillations, which may affect solution accuracy and stability \cite{Colonius2004}. 

In this respect, modern deep learning techniques, especially Physics-Informed Neural Networks  \cite{Lagaris, raissi2019physics} (PINNs), have emerged as serious contenders in the domain of computational physics. These approaches are progressively challenging the long-standing dominance of classical numerical methods across a wide range of physical applications, such as fluid and solid mechanics \cite{sharma2023review, faroughi2024physics}, quantum optics \cite{ji2023}, black-hole spectroscopy \cite{2023PhRvD.107f4025L}, or radiative transfer \cite{2021JQSRT.27007705M}, among many others. A wealth of literature is available for readers interested in exploring the topic in greater depth, ranging from introductory materials to cutting-edge developments
\cite{cuomo2022scientific,ai5030074,toscano2025pinns}
with a variety of architectures such as Deep Operator Networks\cite{deeponet} (DeepONets), Fourier Neural Operators\cite{li2020fno} (FNOs), and Kolgomorov-Arnold Networks\cite{liu2024kan} (KANs).

Although neural networks naturally represent continuous functions thanks to the \emph{universal approximation theorem}\cite{cybenko1989,HORNIK1989,Pinkus_1999}, some studies\cite{liu2024discontinuity, CAI2021110514,doi:10.1137/23M1568107} have shown that PINNs can also predict discontinuous solutions for linear systems of PDEs. One of the first attempts to solve the nonlinear equations of one-dimensional hydrodynamics using PINNs is attributed to Michoski \emph{et al.} \cite{MICHOSKI2020193}. They observed that the training process often produced unstable solutions unless a viscosity term was included in the equations, corresponding to a Reynolds number on the order of $\sim 300$. A series of subsequent works tackled the instability issue in PINNs by introducing various modifications to the loss function. For example, Patel \emph{et al.} \cite{PATEL2022110754} employed the integral form of the PDEs to define the residuals and included additional penalty terms based on entropy inequalities and the total variation diminishing (TVD) property to enhance stability. Ferrer-Sánchez \emph{et al.} \cite{ferrer2024gradient} identified shock locations as the primary source of instability and proposed the Gradient-Annihilated PINNs (GA-PINNs), in which the residuals are multiplied by an annihilation factor that reduces the influence of these regions in the loss function. This approach results in sharper transitions across contact and shock waves, though it may lead to violations of the jump conditions. Liu \emph{et al.} \cite{liu2024discontinuity} further refined the method by incorporating penalty terms in the loss function to enforce the Rankine–Hugoniot conditions and global conservation, although slight but noticeable discrepancies from the exact solutions in the 1D Riemann problems persist. 

In \cite{LLPINNs}, the authors proposed \textit{locally-linearized physics-informed neural networks} (LLPINNs hereafter) to improve the ability of neural networks to detect shock formation. LLPINNs enhance the accuracy of predictions for 1D Riemann problems by applying a localized adjustment to the Jacobian matrix, restricted to regions where shock formation is expected to occur. 
While the original LLPINN approach was promising, its applicability was limited because neither the shock velocity nor the associated eigenvalue is known
\textit{a priori} in real-world problems. In this study, 
we propose a generalized version of LLPINNs
that calculates of the shock speeds based on neighboring states, ensuring that the Rankine–Hugoniot relations are satisfied by design. This approach enables the neural network to dynamically solve the governing equations without needing prior knowledge of the target solution.

Modifying the Jacobian matrix of a system of conservation laws is not a novel idea—it is, in fact, a cornerstone of some Riemann approximate solvers such as the one proposed by Roe\cite{Roe1997}. Building upon the concepts introduced in LLPINNs and the Roe solver, we propose the \emph{locally-Roe PINNs} (LRPINNs). In this approach, the Roe-averaged Jacobian matrix is applied exclusively in regions identified as containing shock waves, while the original formulation remains unchanged elsewhere in the domain. This selective application ensures accurate enforcement of the jump conditions across discontinuities. Furthermore, we extend our generalized LLPINNs and the LRPINNs to two-dimensional problems by incorporating these Jacobian modifications independently along each coordinate direction, a strategy commonly referred to as \emph{dimensional splitting} \cite{toro2009riemann}.

This work is organized as follows. Section II provides a brief introduction to the basic principles of PINNs. In Section III, we outline the key concepts underlying shock formation in hyperbolic systems of conservation laws and present the specific system of equations that is the focus of this study. Section IV presents our results, including one- and two-dimensional examples involving discontinuous solutions, and examines the performance of various approaches. Finally, Section V summarizes our main findings and conclusions.

\section{Physics-informed neural networks}
\subsection{Fundamentals of PINNs}

In this section, we concisely introduce the fundamental concepts underlying PINNs. 
Let $\mathbf{q} \in \mathbb{R}^m$ denote the solution of a time-dependent PDE we aim to approximate in a time interval 
$\left[ 0, T \right]$ using the neural network. Assume it satisfies the following initial-boundary value problem
\begin{align} 
&\mathcal{D}\left[\mathbf{q}\right]\left(\mathbf{x},t\right) = 0, \quad \left(\mathbf{x},t\right) \in \Omega \times \left[ 0, T \right], \\
&\mathbf{q}\left(\mathbf{x},t=0 \right) = \mathbf{q}_0(\mathbf{x}), \quad \mathbf{x} \in \Omega, \\
&\mathbf{q}(\mathbf{x},t) = \mathbf{f}(\mathbf{x},t), \quad \left(\mathbf{x},t\right) \in \partial\Omega \times \left[ 0, T \right],
\end{align}
where $\mathcal{D}$ is a non-linear differential operator and $\Omega$ denotes the spatial domain with boundary $\partial \Omega$. The PINN algorithm yields an approximate solution, $\mathbf{\hat{q}} = \mathbf{\hat{q}}(\mathbf{x}, t; \Theta)$, by constructing a neural network surrogate whose accuracy depends on a set of parameters $\Theta$. These parameters are determined iteratively by minimizing the discrepancy between the predicted 
$\mathbf{\hat{q}}$ and the true solution $\mathbf{q}$. To this end, we introduce the following sets 
\begin{align}
    &\Omega^{(D)} =\left \lbrace \mathbf{x}_i, t_i \right \rbrace_{i=1}^{N_R} \subset \Omega \times (0,T), \label{set:OmegaD} \\
    & \Omega^{(I)} = \left \lbrace \mathbf{x}_i, t_i=0 \right \rbrace_{i=1}^{N_I} \subset \Omega \times \left \lbrace 0 \right \rbrace, \label{set:omegaI} \\
    & \Omega^{(B)} = \left \lbrace \mathbf{x}_i, t_i \right \rbrace_{i=1}^{N_B} \subset \partial\Omega \times (0,T), \label{set:omegaB}
\end{align}
and we minimize the \emph{loss} function
\begin{equation}\label{eq:loss}
    \mathcal{J}\left(\Theta \right) = \mathbf{\omega} \cdot \mathbf{\mathcal{L}}
\end{equation}
with $\mathbf{\omega} \equiv (\omega_D,\omega_I,\omega_B)$ being a vector of user-defined non-negative parameters and $\mathbf{\mathcal{L}} \equiv  (\mathcal{L}_D,\mathcal{L}_I,\mathcal{L}_B)$, where
\begin{align}
    &\mathcal{L}_D = \frac{1}{N_R}\sum_{\left(\mathbf{x},t \right) \in \Omega^{(D)}}{\left[\mathcal{D}\left[\mathbf{\hat{q}}\right]\left(\mathbf{x},t; \Theta\right) \right]^2}, \label{eq:Ld} \\
    &\mathcal{L}_I = \frac{1}{N_I}\sum_{\left(\mathbf{x},t=0 \right) \in \Omega^{(I)}}\left[\mathbf{\hat{q}}\left(\mathbf{x},t=0;\Theta \right) - \mathbf{q}_0(\mathbf{x}) \right]^2, \label{eq:Li} \\
    &\mathcal{L}_B = \frac{1}{N_B}\sum_{\left(\mathbf{x},t \right) \in \Omega^{(B)}}{\left[\mathbf{\hat{q}}\left(\mathbf{x},t;\Theta \right) - \mathbf{f}\left(\mathbf{x},t \right) \right]^2}, \label{eq:Lb}
\end{align}
are, respectively, the mean-squared error of the PDE, initial, and boundary residuals. 

Additional physical constraints (e.g. positivity preservation of some quantities) can also potentially be incorporated as extra terms in the loss function. One of the main advantages of PINNs lies precisely in their inherent flexibility to integrate physical constraints directly into the training process. A more systematic investigation into how such constraints can be incorporated into the PINN framework is certainly justified, and we plan to explore this direction in future work.

For short evolution times, where the fastest wave does not reach the domain boundaries, the results are independent of the boundary conditions and the specific value of $\omega_B$. 
Consequently, for all tests discussed in this paper, we set $\omega_B=0$, leaving the spatial boundaries unconstrained. However, this approach is not suitable for general problems, such as traveling waves in periodic domains over long timescales, where appropriate boundary conditions must be carefully applied.

To minimize the loss function \eqref{eq:loss}, it is necessary to compute its gradient with respect to the parameter set $\Theta$, as well as the derivatives of $\mathbf{\hat{q}}$ with respect to the input variables $\left(\mathbf{x},t\right)$. This is accomplished using Automatic Differentiation (AD), which exploits the fact that the neural network output, 
regardless of its complexity, is constructed through a sequence of arithmetic operations and elementary functions\cite{JMLR:v18:17-468, PMID:37346667,abadi2016tensorflow}. AD enables the efficient transformation of the original initial-boundary problem into a tractable optimization task. 
In the following subsection, we provide a brief overview of the selected optimization method. We refer the interested reader to \cite{toscano2025pinns} for a comprehensive review and an extensive list of references on optimization strategies.

\subsection{Optimization algorithm}

The selection of the optimization algorithm plays a crucial role not only in deep learning applications but also in the broader context of numerical methods and scientific computing. Within the domain of neural networks, the deep learning community has predominantly relied on \emph{first-order} optimization methods—that is, algorithms that use only gradient information of the loss function—due to their computational efficiency and modest memory requirements \cite{ruder2016overview,Soydaner,abdulkadirov2023survey}. This preference is largely motivated by the scale of modern neural network models, which often involve millions of parameters \cite{he2016deep,NIPS2012_c399862d,devlin2019bert}, all of which must be tuned to effectively capture the underlying patterns in complex datasets.

Among the different first-order optimization methods, the Adam\cite{Adam} optimization algorithm is one of the most popular choices \cite{pmlr-v119-alacaoglu20b,Mohamed2023,schmidt2021descending,app10217817}. The Adam optimizer is widely used in many PINN applications (see, for example \cite{WANG2024116813, SAPINNs, ANAGNOSTOPOULOS2024116805, wang2024piratenets}, and references therein) for several reasons: it does not require an extensive search of hyperparameters\cite{Adam,schmidt2021descending}, it is robust on noisy and sparse data\cite{Adam}, and usually converges faster than its main competitors\cite{Adam,Soydaner,schmidt2021descending}, such as Stochastic Gradient Descent \cite{robbins1951stochastic} (SGD), RMSprop\cite{hinton2012rmsprop} or AdaGrad\cite{Adagrad}.

However, it is becoming increasingly clear that improving both the accuracy and computational efficiency of PINNs benefits from the use of \emph{second-order} optimization methods, which exploit information from the Hessian matrix. 
These methods have demonstrated substantial gains in performance across a broader range of problems \cite{urban2025unveiling, kiyani2025optimizer, wang2025gradient}.
Among these, quasi-Newton algorithms 
are effective optimization methods, valued for their robust theoretical basis and good convergence rates in many PINNs applications\cite{urban2025unveiling,kiyani2025optimizer}.
Below, we briefly describe how quasi-Newton methods work (further details are available in
 \cite{urban2025unveiling,kiyani2025optimizer}). 
 
 Let $\Theta_k$ denote the set of trainable variables at optimization step $k$. The update follows the rule
\begin{equation}
    \Theta_{k+1} = \Theta_k - \alpha_k \mathbf{H}_k \nabla \mathcal{J}(\Theta_k),
\end{equation}
where $\alpha_k \in \mathbb{R}$ is the step-length, and $\mathbf{H}_k$ is a symmetric, positive-definite matrix which satisfies the \emph{secant condition}
\begin{equation}
    \mathbf{H}_{k+1}\mathbf{y}_k = \mathbf{s}_k,
\end{equation}
where
\begin{align}
    &\mathbf{s}_k \equiv \Theta_{k+1} - \Theta_k, \\
    &\mathbf{y}_k \equiv \nabla\mathcal{J}(\Theta_{k+1}) - \nabla \mathcal{J}(\Theta_k).
\end{align}
There are several formulas proposed to update the matrix $\mathbf{H}_k$ after each iteration. A general class of quasi-Newton methods is given by the \emph{self-scaled} Broyden formula \cite{urban2025unveiling}
\begin{equation}
    \mathbf{H}_{k+1} = \frac{1}{\tau_k}\left[\mathbf{H}_k - \frac{\mathbf{H}_k \mathbf{y}_k\mathbf{y}_k^T \mathbf{H_k}}{\mathbf{y}_k^T \mathbf{H}_k \mathbf{y}_k} + \phi_k \mathbf{v}_k \mathbf{v}_k^T\right] + \frac{\mathbf{s}_k \mathbf{s}_k^T}{\mathbf{y_k}^T \mathbf{s}_k}, 
\end{equation}
with $\tau_k$ and $\phi_k$ being real scalars that may vary between iterations, and
\begin{equation}
    \mathbf{v}_k = \sqrt{\mathbf{y}_k^T \mathbf{H}_k \mathbf{y}_k}\left[\frac{\mathbf{s}_k}{\mathbf{y}_k^T \mathbf{s_k}} - \frac{\mathbf{H_k \mathbf{y_k}}}{\mathbf{y}_k^T \mathbf{H}_k \mathbf{y}_k} \right].
\end{equation}
Particular choices of $\tau_k$ and $\phi_k$ are associated with different quasi-Newton algorithms. For example, for $\tau_k=1$ and $\phi_k=1$ we recover the so-called BFGS method \cite{BroydenBFGS,FletcherBFGS,Goldfarb1970AFO,ShannoBFGS}. These choices are not arbitrary, but should respect certain conditions to ensure \emph{superlinear} convergence \cite{al2005wide,Al-BaaliSpedicatoMaggioni,ContrerasTapia}. For this paper, we have choose the combination suggested in \cite{ContrerasTapia}
\begin{align}
    &\tau_k = \min \left \lbrace 1,\frac{1}{b_k} \right \rbrace, \\
    &\phi_k = 0,
\end{align}
where 
\begin{equation}
    b_k = -\alpha_k \frac{\mathbf{s}_k^T\nabla \mathcal{J}(\Theta_k)}{\mathbf{y}_k^T \mathbf{s}_k}.
\end{equation}
We found that this combination yields faster convergence rates than the \emph{SSBFGS} algorithm and similar rates to the \emph{SSBroyden} algorithm, both described in \cite{urban2025unveiling}. However, we also observed that the use of \emph{SSBroyden} may lead to numerical errors, as it involves expressions with square roots, which can affect the stability of the optimization process in certain cases. Finally, for the line search to determine $\alpha_k$ at each iteration, we employed the backtracking method detailed in \cite{dennis1996numerical}.

\section{Hyperbolic conservation laws}
\subsection{Fundamental concepts}

We begin this section by considering the one-dimensional system of conservation laws in its differential form:
\begin{equation}\label{eq:conservative1D}
    \frac{\partial \mathbf{U}}{\partial t} + \frac{\partial \mathbf{F}\left(\mathbf{U}\right)}{\partial x} = 0,
\end{equation}
where $\mathbf{U}$ represents the vector of conserved quantities, and  $\mathbf{F}\left(\mathbf{U}\right)$ is the vector of fluxes, with each component depending on the components of $\mathbf{U}$. If we assume that the flux function $\mathbf{F}$ is differentiable, we can rewrite the system \eqref{eq:conservative1D} in quasi-linear form
\begin{equation}\label{eq:quasilinear1D}
    \frac{\partial \mathbf{U}}{\partial t} + \mathbf{A}\left(\mathbf{U} \right)\frac{\partial \mathbf{U}}{\partial x} = 0,
\end{equation}
where $\mathbf{A} \equiv \frac{\partial \mathbf{F}}{\partial \mathbf{U}}$ is the Jacobian of $\mathbf{F}$.
In this paper, we restrict ourselves to the case of \emph{hyperbolic} systems, where $\mathbf{A}$ is diagonalizable with real eigenvalues $\lbrace \lambda_i \rbrace_{i=1}^m$. 

If $\mathbf{A}$ is not constant, equations \eqref{eq:conservative1D} or \eqref{eq:quasilinear1D} represent a non-linear system of conservation laws. The characteristic curves associated with the $i$th eigenvalue, $\frac{dx}{dt} = \lambda_i$, may then intersect, leading to the formation of a discontinuity \cite{leveque1998computational,toro2009riemann} (a so-called \emph{shock wave} in the case of hydrodynamics). 
When this occurs, the solution becomes discontinuous, and we cannot express the conservation laws in the differential forms given by \eqref{eq:conservative1D} or \eqref{eq:quasilinear1D}. Instead, the solution $\mathbf{U}$ must satisfy the integral form of these equations (weak solutions), obtained by multiplying \eqref{eq:conservative1D} by a smooth function $\phi$ with compact support and then integrating over the domain. This yields \cite{toro2009riemann}
\begin{equation*}\label{eq:weak_solution}
    \int_{\left[0,+\infty \right) \times \mathbb{R}}\left[\frac{\partial \phi}{\partial t}\mathbf{U} + \frac{\partial \phi }{\partial x} \mathbf{F}\left(\mathbf{U}\right)\right] \text{d}x\text{d}t = -\int_{\mathbb{R}}{\phi(x,0)\textbf{U}(x,0) \text{d}x}.
\end{equation*}
From this identity, it can be shown that if $\mathbf{U}$ is differentiable except at the curve associated with the crossing of the $\lambda_i$ characteristic, then the values of $\mathbf{U}$ on both sides of this curve are related by
\begin{equation}\label{eq:RH}
    \mathbf{F}\left(\mathbf{U}_R\right) - \mathbf{F}\left(\mathbf{U}_L\right) = s_i \left(\mathbf{U}_R - \mathbf{U}_L\right),
\end{equation}
which are the so-called Rankine-Hugoniot conditions \cite{rankine1870, hugoniot1887, hugoniot1889}. Here, the subindices $L$ and $R$ denote the corresponding values of $\mathbf{U}$ evaluated at the left and right sides of the shock, while $s_i$ is the velocity at which this discontinuity moves. 

Given the states $\mathbf{U}_R$ and $\mathbf{U}_L$, the Rankine-Hugoniot conditions \eqref{eq:RH} can be thought of as a set of algebraic equations that are fulfilled if the velocity of the discontinuity is equal to $s_i$. However, the solution of this system of equations is not unique in general. To exclude non-physical solutions, a simple condition due to \cite{Lax1957entropy} states that
\begin{eqnarray}\label{eq:lax_entropy}
    \lambda_{i-1} \left(\mathbf{U}_L \right) < s_i < \lambda_i \left(\mathbf{U}_L \right), 
    \nonumber \\
    \lambda_i \left(\mathbf{U}_R \right) < s_i < \lambda_{i+1} \left(\mathbf{U}_R \right)
\end{eqnarray}
provided that $\lambda_i$ is a genuinely nonlinear field.
That is, $\lambda_i$ satisfies the inequality $\nabla_{\mathbf{U}} \lambda \left(\mathbf{U} \right) \cdot \mathbf{R}^{(i)}\left(\mathbf{U} \right) \neq 0$ for all $\mathbf{U}$, where $\mathbf{R}^{(i)}$ is the right eigenvector associated with the eigenvalue $\lambda_i$ and $\nabla_{\mathbf{U}}$ denotes the gradient with respect to the $\mathbf{U}$ components\cite{leveque1998computational,toro2009riemann}.

These basic concepts are the basis of  HRSC methods. These numerical methods accurately model discontinuities like shock waves and contact surfaces, achieving high accuracy in smooth regions while robustly handling sharp gradients without introducing spurious oscillations. The Riemann problem, which solves the evolution of two constant states separated by a discontinuity, is central to their design. It provides the exact or approximate solution to local wave interactions at cell interfaces, enabling HRSC methods to ensure stability and accuracy. In this work, we will apply all these fundamental concepts to improve PINNs performance.

\subsection{Important physical applications}

\subsubsection{Euler equations}

We begin with a well-known example: the Euler equations, a nonlinear system of conservation laws that describe the evolution of an adiabatic, inviscid flow
\cite{toro2009riemann,leveque1998computational}.
The Euler equations in the form of \eqref{eq:conservative1D} are:
\begin{equation}
    \mathbf{U} =
    \begin{pmatrix}
    \rho \\
    \rho u \\
    E
    \end{pmatrix},
    \quad 
    \mathbf{F\left(\mathbf{U} \right)}=
    \begin{pmatrix}
        \rho u \\
        \rho u^2 + p \\
        u\left(E + p \right)
    \end{pmatrix},
\end{equation}
where $\rho$ is the mass density, $u$ the fluid velocity, and $E$ is the total energy density
\begin{equation*}
    E = \frac{\rho u^2}{2} + \rho\varepsilon,
\end{equation*}
where $\varepsilon$ is the specific internal energy. 
These three equations represent the fundamental laws of the conservation of mass, momentum, and total energy. 
The pressure $p$ is provided by an equation of state $p = p \left(\rho, \varepsilon \right)$ that closes the system. 
A classical example is the ideal adiabatic gas
\begin{equation}\label{eq:eos_euler}
    p = \left(\gamma-1\right) \rho \varepsilon,
\end{equation}
where $\gamma > 0$ is the adiabatic index. 

The Euler equations can also be expressed in the quasi-linear form defined in \eqref{eq:quasilinear1D}, with
\begin{equation}\label{eq:jacobian_euler}
\mathbf{A} = 
\begin{bmatrix}
0 & 1 & 0 \\
\frac{\gamma-3}{2}u^2 & -\left(\gamma-3 \right) u & \gamma -1 \\
u \left(-H + \frac{\gamma -1}{2}u^2 \right) & H - \left(\gamma -1 \right)u^2 & \gamma u
\end{bmatrix},
\end{equation}
where $H$ is the total specific enthalpy
\begin{equation}\label{eq:specific_enthalpy}
    H = \frac{E + p}{\rho} = \frac{a^2}{\gamma-1} + \frac{u^2}{2},
\end{equation}
and $a$ is the sound speed
\begin{equation}\label{eq:a_euler}
    a = \sqrt{\frac{\gamma p}{\rho}}.
\end{equation}
The Jacobian for the Euler equations is diagonalizable with real eigenvalues
\begin{equation}
    \left(\lambda_-, \lambda_0,\lambda_+ \right) = \left(u -a,u,u+a \right),
\end{equation}
and the corresponding right eigenvectors
\begin{equation}
    \mathbf{r}^{(-)} = 
    \begin{bmatrix}
        1 \\
        u-a \\
        H-ua
    \end{bmatrix},
    \mathbf{r}^{(0)}=
    \begin{bmatrix}
        1 \\
        u \\
        u^2/2
    \end{bmatrix},
    \mathbf{r}^{(+)} = 
    \begin{bmatrix}
       1 \\
       u+a \\
       H + ua
    \end{bmatrix}.
\end{equation}

The formation of shock waves in the Euler equations is due to the crossing of the characteristics associated with the eigenvalues $\lambda_-$ and $\lambda_+$. Applying the Rankine-Hugoniot conditions together with the Lax entropy conditions, the shock speeds associated with these eigenvalues are\cite{toro2009riemann}
\begin{align}
    s_- = u_L - a_L \sqrt{\frac{1}{2\gamma}\left[\left(\gamma+1 \right) \frac{p_R}{p_L} + \gamma-1 \right]}, \label{eq:s1_euler}\\
    s_+ = u_R + a_R \sqrt{\frac{1}{2\gamma}\left[\left(\gamma+1 \right) \frac{p_L}{p_R} + \gamma-1 \right]} \label{eq:s3_euler},
\end{align}
Note that, in the limit of a \emph{weak} shock wave, these expressions tend, respectively, to $\lambda_-$ and $\lambda_+$.

Finally, associated with the eigenvalue $\lambda_0$, we have a \emph{contact} wave, characterized by a discontinuity in the density but not in the velocity or pressure. In this case, the characteristic curves do not intersect, but instead are parallel to this wave and, consequently, the discontinuity moves with velocity $s_0 = \lambda_0 = u$.

\subsubsection{Special relativistic hydrodynamics}

We also study the evolution of a relativistic fluid in special relativity. The resulting equations again form a system of conservation laws, and they arise from the conservation of baryon number and the conservation of energy-momentum
\begin{align}
    &\mathbf{\nabla} \cdot \mathbf{J} = 0, \label{eq:conservationJ}\\
    &\mathbf{\nabla} \cdot \mathbf{T} = 0 \label{eq:_conservationT},
\end{align}
where $\mathbf{J}$ and $\mathbf{T}$ are respectively the rest-mass current and the energy-momentum tensor. If we consider a perfect fluid, their components are\footnote{In this case, we work in $c=1$ units}
\begin{align}
    J^\mu &= \rho u^\mu, \label{eq:J}\\ 
    T^{\mu \nu} &= \rho h u^\mu u^\nu + p \eta^{\mu \nu}, \label{eq:T}
\end{align}
where $\rho$ is the rest-mass density, $p$ is the pressure, $h=1+\varepsilon + p/\rho$ is the specific enthalpy, with $\varepsilon$ being the specific internal energy, $\eta^{\mu \nu} = \text{diag}(-1,1,1,1)$ is the Minkowski metric, and $u^\mu$ is the four-velocity of the fluid. 

Substituting \eqref{eq:J} and \eqref{eq:T} into the general form of the conservation equations \eqref{eq:conservationJ} and \eqref{eq:_conservationT}, we obtain again a system of conservation laws in differential form, with
\begin{equation}
    \mathbf{U} =
    \begin{pmatrix}
    \rho W\\
    \rho h W^2u \\
    \rho h W^2 - p
    \end{pmatrix},
    \quad 
    \mathbf{F\left(\mathbf{U} \right)}=
    \begin{pmatrix}
        \rho Wu \\
        \rho h W^2 u^2 + p  \\
         \rho h W^2u
    \end{pmatrix},
\end{equation}
where $W = \left(1-u^2 \right)^{-1/2}$ is the Lorentz factor. We observe that the introduction of the Lorentz factor and the specific enthalpy makes the dependence between the fluxes and the conserved variables much more complicated than in the non-relativistic case. 

The Jacobian matrix $\mathbf{A} \equiv \frac{\partial \mathbf{F}}{\partial \mathbf{U}}$ can be expressed as
\begin{equation}
    \mathbf{A} = \mathbf{A}_1 \mathbf{A}_0^{-1},
\end{equation}
where $\mathbf{A}_0 \equiv \frac{\partial \mathbf{U}}{\partial \mathbf{w}}$ and $\mathbf{A}_1 \equiv \frac{\partial \mathbf{F}}{\partial \mathbf{w}}$ are respectively the Jacobians of the conserved variables and the fluxes with respect the vector of primitive variables $\mathbf{w} = \left(\rho,u,\varepsilon \right)$. From here, one can find the eigenvalues $\lambda$ and right eigenvectors $\mathbf{f}$ of the Jacobian $\mathbf{A}$ by solving
\begin{equation*}
    \left(\mathbf{A}_1 - \lambda \mathbf{A}_0 \right)\mathbf{v} = 0, 
\end{equation*}
where $\mathbf{v} = \mathbf{A}_0^{-1} \mathbf{r}$. The eigenvalues are
\begin{equation}
    \left(\lambda_-, \lambda_0,\lambda_+ \right) = \left(\frac{u-a}{1-ua},u,\frac{u+a}{1+ua} \right),
\end{equation}
with $a$ being the relativistic sound speed
\begin{equation}
    a = \sqrt{\frac{\gamma p}{h\rho}},
\end{equation}
where we consider that the equation of state $p = p(\rho,\varepsilon)$ is given by \eqref{eq:eos_euler}. The corresponding right eigenvectors are 
\begin{equation}
    \mathbf{r}^{(-)} = 
    \begin{bmatrix}
        1 \\
        hW(u-a) \\
        hW(1-ua)
    \end{bmatrix},
    \mathbf{r}^{(0)}=
    \begin{bmatrix}
        1/W \\
        u \\
        1
    \end{bmatrix},
    \mathbf{r}^{(+)} = 
    \begin{bmatrix}
        1 \\
        hW(u+a) \\
        hW(1+ua)
    \end{bmatrix},
\end{equation}

As in the non-relativistic case, we associate shocks with the eigenvalues $\lambda_-$ and $\lambda_+$, whereas for $\lambda_0$ we obtain a contact discontinuity. Applying the Rankine-Hugoniot and the Lax entropy conditions, the shock speeds in this case are \cite{MartiMuller}
\begin{align}
    s_- = \frac{\rho_L W_L^2 u_L -j^2\sqrt{1+\frac{\rho_L^2}{j^2}}}{\rho_LW_L^2 + j^2}, \\
    s_+ = \frac{\rho_R W_R^2 u_R + j^2\sqrt{1+\frac{\rho_R^2}{j^2}}}{\rho_RW_R^2 + j^2},
\end{align}
where $j$ is the mass flux across the shock
\begin{equation}
    j = \sqrt{- \frac{p_R - p_L}{\frac{h_R}{\rho_R} - \frac{h_L}{\rho_L}}}.
\end{equation}

\subsection{Jacobian modifications}

To incorporate jump conditions across shocks into the training process, we follow a strategy similar to that used in many approximate Riemann solvers by modifying the original system of conservation laws \eqref{eq:conservative1D}. Specifically, we consider the following system of PDEs
\begin{equation}\label{eq:modified_quasilinear1D}
    \frac{\partial \mathbf{U}}{\partial t} + \mathbf{\tilde{A}}\left(\mathbf{U} \right)\frac{\partial \mathbf{U}}{\partial x} = 0,
\end{equation}
with $\mathbf{\tilde{A}}$ being an appropriate modification of the original Jacobian $\mathbf{A}$ which properly encodes the jump conditions across a shock. Hence, the points $(x,t) \in \Omega^{(D)}$ where we calculate the residuals are forced to satisfy the system \eqref{eq:modified_quasilinear1D}.

In \cite{LLPINNs}, they proposed the \emph{locally linearized physics-informed neural networks} (LLPINNs) as a means to improve the predictive capabilities of PINNs for shock formation in hyperbolic systems of conservation laws. To achieve this, they embed information about shock propagation into the Jacobian matrix by selectively modifying it in regions where shock formation is likely to occur. In this modification, they alter only the diagonal matrix of eigenvalues, replacing the eigenvalue associated with shock formation with the corresponding shock velocity.

The two main steps in this approach are:
\begin{enumerate}
\item {\bf Localize a possible discontinuity.}
Define the potential region of shock formation as the set of points where we detect compression, that is
\begin{equation}\label{eq:shock_criterion}
    \Omega_s = \left \lbrace (x,t) \in \Omega^{(D)} \mid \frac{\partial u}{\partial x} \leq -M   \right \rbrace,
\end{equation}
with $u$ being the fluid velocity and $M$ is a positive constant. Although shocks are characterized by large negative values of $\partial u/\partial x$, in practice it is preferable to consider a relatively small value of $M$ to ensure that we have enough points in $\Omega_s$ for the Jacobian modification to act.

\item Only in such regions, {\bf use a suitable modified Jacobian} that captures features of discontinuous solutions.
The modified Jacobian matrix is
\begin{equation}\label{eq:jacobian_LLPINNs}
    \mathbf{\tilde{A}} = \mathbf{R} \tilde{\Lambda}\mathbf{R}^{-1},
\end{equation}
where $\textbf{R}$ is the matrix of right eigenvectors and, if the shock is associated with the $i$th eigenvalue, then
\begin{equation}\label{eq:tilde_lambda}
    \tilde{\Lambda} = \text{diag}\left(\lambda_1,\lambda_2, \ldots, \tilde{\lambda}_i, \ldots, \lambda_n \right),
\end{equation}
with $\tilde{\lambda}_i$ defined as
\begin{equation}\label{eq:replacement_llpinns}
    \tilde{\lambda}_i=
    \begin{cases}
        s_i, \quad (x,t) \in \Omega_s, \\
        \lambda_i, \quad \text{otherwise},
    \end{cases}
\end{equation}
where $s_i$ denotes the shock speed determined by the Rankine–Hugoniot conditions. 
\end{enumerate}

Although simple, this idea elegantly incorporates information from jump conditions, enabling the neural network to capture shock formation more accurately. However, we identified two main limitations in their methodology: (1) the eigenvalue associated with the formation of the shock was replaced by the exact numerical value of the corresponding shock speed\footnote{An exception is made for the Burgers equation, where the shock speed is calculated based on the left and right states of the velocity.}, which is not known a priori and is problem dependent; and (2) the selection of the eigenvalue to be modified was itself not determined in advance.

To address these two limitations, we propose adding two extra steps before modifying the Jacobian matrix:
\begin{enumerate}
    \item First, at the points where a shock is detected, we compute the shock speeds associated with all eigenvalues that could give rise to a shock. For example, in the case of the Euler equations, we calculate the shock speeds $s_-$ and $s_+$ corresponding to the eigenvalues $\lambda_-$ and $\lambda_+$, respectively, using formulas \eqref{eq:s1_euler} and \eqref{eq:s3_euler}. We remark that while $\lambda_i$ depends solely on the function values predicted by the PINN at the point $\boldsymbol{x}_s$, computing the wave speeds $s_i$ requires information from neighboring points on both sides of a given shock point $\boldsymbol{x}_s$. Hence, to compute the wave speeds $s_1$ and $s_3$ at $\boldsymbol{x}_s = (t_s, x_s)$, the left state $(\rho_L, u_L, p_L)$ and the right state $(\rho_R, u_R, p_R)$ are defined by evaluating the neural network at the neighboring points $(t_s, x_s - \Delta x)$ and $(t_s, x_s + \Delta x)$, respectively, where $\Delta x$ is a hyperparameter.
    \item We apply the Lax entropy conditions \eqref{eq:lax_entropy} to determine which eigenvalue should be modified in \eqref{eq:tilde_lambda}.
\end{enumerate}

It is worth noting that, although the previous discussion focused on a single shock wave, the approach naturally extends to scenarios involving multiple shock waves by substituting each eigenvalue associated with a shock with its corresponding shock velocity. The whole process, including the extra steps mentioned before, is done at each step of the training process of the PINN.
Notably, outside shock regions, the eigenvalues are unaltered, leaving the Jacobian matrix unchanged.
The resulting algorithm (particularized for the Euler equations for illustration purposes) is described below in Algorithm \ref{alg:LLPINNs}.

\begin{figure}
\begin{algorithm}[H]
\caption{Generalized LLPINNs for Euler equations}\label{alg:LLPINNs}
\begin{algorithmic}[1]
  \State \textbf{Input:} $\Omega^{(D)}$, $\Omega^{(I)}$, $\Omega^{(B)}$, $M$, $\Delta x$, $\omega \equiv(\omega_D,\omega_I,\omega_B)$, $\eta$.
  \State \textbf{Output:} PINN solution $\mathbf{U}\left(x,t; \Theta\right)$
  \State Initialize trainable variables $\Theta$
  \While{$\mathcal{J}(\Theta) > \eta$}
  \State Calculate $\partial_x u_ \equiv \frac{\partial u (\mathbf{x})}{\partial x}$ for each point $\mathbf{x} \in \Omega^{(D)}$.
  \State Determine the points $\mathbf{x}_s = \left(x_s, t_s \right) \in \Omega_s$ using the criterion \eqref{eq:shock_criterion}.
  \State Compute $\mathbf{x}_{L} = (x_s-\Delta x,t_s)$ and $\mathbf{x}_{R} = (x_s+\Delta x,t_s)$.
  \State Obtain left and right states $\mathbf{U}_L = \mathbf{U}\left(\mathbf{x}_L;\Theta \right)$, $\mathbf{U}_R = \mathbf{U}\left(\mathbf{x}_R;\Theta \right)$.
  \State Compute the possible shock velocities $s_-$ and $s_+$, given by the expressions \eqref{eq:s1_euler} and \eqref{eq:s3_euler}.
  \State Use Lax entropy conditions \eqref{eq:lax_entropy} to determine if $\textbf{x}_s$ belongs to the $\lambda_-$ or to the $\lambda_+$ shock.
  \State Change $\lambda_-$ or $\lambda_+$ by $s_-$ and $s_+$ respectively, according to the result obtained in the previous step.
  \State Compute the modified Jacobian $\tilde{\mathbf{A}}$ given by \eqref{eq:jacobian_LLPINNs}.
  \State Calculate $\mathcal{L}_D$, defined in \eqref{eq:Ld}, with the PDE residuals $\mathcal{D}\left[\mathbf{U} \right](x,t) = \frac{\partial \mathbf{U}}{\partial t} + \mathbf{\tilde{A}}\left(\mathbf{U} \right)\frac{\partial \mathbf{U}}{\partial x}$ for each point $(x,t) \in \Omega^{(D)}$.
  \State Calculate $\mathcal{L}_I$ and $\mathcal{L}_B$, respectively defined in \eqref{eq:Li} and \eqref{eq:Lb}. for the points $(x,t)$ belonging to $\Omega^{(I)}$ and $\Omega^{(R)}$.
  \State Calculate $\mathcal{J}(\Theta)$ using \eqref{eq:loss} and perform an optimization step.
  \EndWhile
\end{algorithmic}
\end{algorithm}
\end{figure}

Furthermore, the main problem of LLPINNs is that the modified Jacobian, in general, does not satisfy the key property
\begin{equation}\label{eq:conservationA}
    \mathbf{F}\left(\mathbf{U}_R\right)-\mathbf{F}\left(\mathbf{U}_L\right) = \tilde{\mathbf{A}}\left(\mathbf{U}_R-\mathbf{U}_L \right),
\end{equation}
which ensures conservation across discontinuities\cite{Roe1997,toro2009riemann}.As we will show in the corresponding results section, LLPINNs correctly predict the shock velocities but not the jump in the conserved variables, which ultimately leads to a violation of global conservation.

To address this, we incorporate jump conditions by defining
$\mathbf{\tilde{A}}$ to be the {\bf Roe-averaged} Jacobian matrix, originally introduced in \cite{Roe1997}. 
The Roe solver uses the exact solution of the Riemann problem for a constant Jacobian matrix, where the nonlinear system is averaged around a mean state, termed the Roe average.

The resulting Jacobian, $\tilde{\mathbf{A}}$, which now depends on both the left and right states, must have the following properties:
\begin{itemize}
    \item $\tilde{\mathbf{A}}$ must be diagonalizable with real eigenvalues, just like the original Jacobian.
    \item If the left ($\mathbf{U}_L$) and right ($\mathbf{U}_R$) states tend to the same state $\mathbf{U}$, then we must recover the original Jacobian matrix $\mathbf{A}(\mathbf{U})$. 
    \item For any $\mathbf{U}_L$, $\mathbf{U}_R$, equation \eqref{eq:conservationA} should be satisfied. This condition ensures that, in the presence of a single discontinuity at the interface, the linearized problem provides a solution that matches the full non-linear Riemann problem.
\end{itemize}
Note in particular that the Jacobian of LLPINNs fulfills the first two conditions, but not the third one. 

For the Euler equations, the Roe Jacobian matrix is given again by the matrix \eqref{eq:jacobian_euler}, but we replace the velocity and the total specific enthalpy by the following averages
\begin{align}
    \tilde{u} = \frac{\sqrt{\rho_L}u_L + \sqrt{\rho_R}u_R}{\sqrt{\rho_L} + \sqrt{\rho_R}}, \label{eq:u_average} \\
    \tilde{H} = \frac{\sqrt{\rho_L}H_L + \sqrt{\rho_R}H_R}{\sqrt{\rho_L} + \sqrt{\rho_R}}, \label{eq:H_average}
\end{align}
and the averaged sound speed is given by
\begin{equation}
    \tilde{a} = \sqrt{\left(\gamma-1 \right)\left(\tilde{H}-\frac{\tilde{u}^2}{2} \right)}
\end{equation}


The Roe Jacobian for the relativistic case was developed by Eulderink \& Mellema \cite{EulderinkMellema} and is expressed in terms of the averaged state
\begin{equation}
    \tilde{\mathbf{w}} = \frac{\mathbf{w}_L + \mathbf{w_R}}{\sqrt{\rho_Lh_L} + \sqrt{\rho_R h_R}}, 
\end{equation}
where
\begin{align}
    \textbf{w} = \sqrt{\rho h}\left(W,uW, \frac{p}{\rho h}\right).
\end{align}

Apart from the choice of the Jacobian, the methodology in this case follows the same approach as in the LLPINNs. Specifically, we first identify the regions of potential shock formation based on the criterion defined in \eqref{eq:shock_criterion}. Then, we apply the quasi-linear form of the PDEs with the modified Jacobian matrix only at the points within these regions, while using the original Jacobian for the other points. The Algorithm \ref{alg:lrpinns} outlines the methodology for this case. We refer to this approach as LRPINNs (with the ``R'' standing for Roe) to distinguish it from the original LLPINN methodology. 

Both methods (generalized LLPINNs and LRPINNs) rely on non-local information — either the shock speed or the Roe-averaged matrix. Note that the LRPINNs, by design, implicitly incorporate the information of the shock velocities from equation \eqref{eq:conservationA}, thereby superseding Algorithm \ref{alg:LLPINNs} by embedding more physical information into the learning process.

\begin{figure}
\begin{algorithm}[H]
\caption{LRPINNs}\label{alg:lrpinns}
\begin{algorithmic}[1]
\State \textbf{Input:} $\Omega^{(D)}$, $\Omega^{(I)}$, $\Omega^{(B)}$, $M$, $\Delta x$, $\omega \equiv(\omega_D,\omega_I,\omega_B)$, $\eta$.
  \State \textbf{Output:} PINN solution $\mathbf{U}\left(x,t; \Theta\right)$
  \State Initialize trainable variables $\Theta$
  \While{$\mathcal{J}(\Theta) > \eta$}
  \State Calculate $\partial_x u_ \equiv \frac{\partial u (\mathbf{x})}{\partial x}$ for each point $\mathbf{x} \in \Omega^{(D)}$.
  \State Determine the points $\mathbf{x}_s = \left(x_s, t_s \right) \in \Omega_s$ using the criterion \eqref{eq:shock_criterion}.
  \State Compute $\mathbf{x}_{L} = (x_s-\Delta x,t_s)$ and $\mathbf{x}_{R} = (x_s+\Delta x,t_s)$.
  \State Obtain left and right states $\mathbf{U}_L = \mathbf{U}\left(\mathbf{x}_L;\Theta \right)$, $\mathbf{U}_R = \mathbf{U}\left(\mathbf{x}_R;\Theta \right)$.
  \State Calculate the corresponding Roe averages.
  \State Compute the modified Jacobian $\tilde{\mathbf{A}}$ in terms of the Roe averages given at the previous step.
  \State For the rest of the points $\mathbf{x}_o \notin \Omega_s$, $\mathbf{\tilde{A}} \gets \mathbf{A}$.
  \State Calculate $\mathcal{L}_D$, defined in \eqref{eq:Ld}, with the PDE residuals $\mathcal{D}\left[\mathbf{U} \right](x,t) = \frac{\partial \mathbf{U}}{\partial t} + \mathbf{\tilde{A}}\left(\mathbf{U} \right)\frac{\partial \mathbf{U}}{\partial x}$ for each point $(x,t) \in \Omega^{(D)}$.
  \State Calculate $\mathcal{L}_I$ and $\mathcal{L}_B$, respectively defined in \eqref{eq:Li} and \eqref{eq:Lb}. for the points $(x,t)$ belonging to $\Omega^{(I)}$ and $\Omega^{(R)}$.
  \State Calculate $\mathcal{J}(\Theta)$ using \eqref{eq:loss} and perform an optimization step.
  \EndWhile
\end{algorithmic}
\end{algorithm}
\end{figure}

\subsection{Multidimensional Euler equations: Dimensional splitting}

A straightforward approach to extend one-dimensional classical methods to higher dimensions is to apply them along each coordinate direction. This approach is commonly referred to as \emph{dimensional splitting} \cite{toro2009riemann}. Naturally, one may question whether dimensional splitting can also be applied to the LLPINNs and LRPINN algorithms described in the previous sections. In this paper, we explore the 2D Euler equations as a proof of concept.

The 2D Euler equations can be written in the following conservative form
\begin{equation}\label{eq:conservative2D}
    \frac{\partial \mathbf{U}}{\partial t} + \frac{\partial \mathbf{F}\left(\mathbf{U}\right)}{\partial x} + \frac{\partial \mathbf{G}\left(\mathbf{U}\right)}{\partial y}= 0,
\end{equation}
with
\begin{equation}
    \mathbf{U} =
    \begin{pmatrix}
    \rho \\
    \rho u \\
    \rho v \\
    E
    \end{pmatrix},
    \mathbf{F}\left(\mathbf{U}\right) =
    \begin{pmatrix}
        \rho u \\
        \rho u^2 + p \\
        \rho u v \\
        u(E+p)
    \end{pmatrix},
    \mathbf{G}\left(\mathbf{U}\right) =
    \begin{pmatrix}
        \rho v \\
        \rho uv \\
        \rho v^2 + p \\
        v(E+p).
    \end{pmatrix},
\end{equation}
where now $u$ and $v$ are the components of the vector velocity in the $x$ and $y$ directions, respectively. The total energy density is now expressed as
\begin{equation}
    E = \frac{\rho}{2}V^2 + \frac{p}{\gamma-1}.
\end{equation}
Here, $V^2 = u^2 + v^2$. We also assume that the equation of state $p=p(\rho,\varepsilon)$ remains the same as given in \eqref{eq:eos_euler}.

On the other hand, the 2D Euler equations can also be expressed in the following quasi-linear form
\begin{equation}\label{eq:quasilinear2D}
    \frac{\partial \mathbf{U}}{\partial t} + \mathbf{A}\left(\mathbf{U} \right)\frac{\partial \mathbf{U}}{\partial x} + \mathbf{B}\left(\mathbf{U} \right)\frac{\partial \mathbf{U}}{\partial y}= 0,
\end{equation}
where we define the Jacobian matrices
\begin{align}
    &\mathbf{A} = 
\begin{bmatrix}
0 & 1 & 0 & 0\\
-u^2 + \frac{\gamma-1}{2}V^2 & -\left(\gamma-3 \right) u & -(\gamma-1)v & \gamma -1 \\
-uv & v & u & 0 \\
u \left(-H + \frac{\gamma -1}{2}V^2 \right) & H - \left(\gamma -1 \right)u^2 & -(\gamma-1)uv & \gamma u
\end{bmatrix}, \label{eq:A2D}\\
&\mathbf{B} = 
\begin{bmatrix}
0 & 0 & 1 & 0\\
-uv & v & u & 0 \\
-v^2 + \frac{\gamma-1}{2}V^2 & -\left(\gamma-1 \right) u & -(\gamma-3)v & \gamma -1 \\
v \left(-H + \frac{\gamma -1}{2}V^2 \right) &  -(\gamma-1)uv & H - \left(\gamma -1 \right)v^2  & \gamma v
\end{bmatrix},\label{eq:B2D}
\end{align}
with the total specific enthalpy calculated as
\begin{equation}\label{eq:specific_enthalpy_2D}
    H = \frac{E + p}{\rho} = \frac{a^2}{\gamma-1} + \frac{V^2}{2}.
\end{equation}

Both Jacobians are, as in the 1D Euler equations, diagonalizable with real eigenvalues
\begin{align}
    &\left(\lambda_-, \lambda_{01},\lambda_{02},\lambda_+ \right) = \left(u -a,u,u,u+a \right), \\
    &\left(\mu_-, \mu_{02},\mu_{03},\mu_+ \right) = \left(v -a,v,v,v+a \right),
\end{align}
where $\left \lbrace \lambda_i \right \rbrace$ and $\left \lbrace \mu_i \right \rbrace$ are respectively the eigenvalues of $\mathbf{A}$ and $\mathbf{B}$, and $a$ is the sound speed defined in \eqref{eq:a_euler}.

In a dimensional splitting approach to the methods mentioned above, the neural network is trained to solve the following equation
\begin{equation}
    \frac{\partial \mathbf{U}}{\partial t} + \mathbf{\tilde{A}}\left(\mathbf{U} \right)\frac{\partial \mathbf{U}}{\partial x} + \mathbf{\tilde{B}}\left(\mathbf{U} \right)\frac{\partial \mathbf{U}}{\partial y}= 0,
\end{equation}
where $\mathbf{\tilde{A}}$ and $\mathbf{\tilde{B}}$ are appropriate modifications of the true Jacobian matrices. As before, these modifications are applied only in regions of potential shock formation, while the matrices remain unchanged elsewhere. We define again the regions of potential shock formation where we detect compression, which is now computed in terms of the divergence of the velocity field 
\begin{equation}\label{eq:shock_criterion_tot}
    \Omega_s = \left \lbrace (x,y,t) \in \Omega^{(D)} \mid \nabla \cdot \mathbf{u} = \frac{\partial u}{\partial x} + \frac{\partial v}{\partial y} \leq -M   \right \rbrace,
\end{equation}
and then, to know in which direction we should take the neighboring points, we introduce the sets  
\begin{align}
    &\Omega_{sx} = \left \lbrace (x,y,t) \in \Omega_s \mid \frac{\partial u}{\partial x} \leq -M_x   \right \rbrace, \label{eq:xshock_criterion}\\
     &\Omega_{sy} = \left \lbrace (x,y,t) \in \Omega_s \mid \frac{\partial v}{\partial y} \leq -M_y   \right \rbrace, \label{eq:yshock_criterion}
\end{align}
where $M_x$ and $M_y$ are again two positive constants. 

In multi-dimensional scenarios, the use of the full divergence in (\ref{eq:shock_criterion_tot}) is crucial, as 
it captures compressive effects that may indicate potential shock formation across multiple coordinate directions. By employing the divergence, 
we obtain a scalar measure of volumetric compression that reflects the combined influence of gradients in all directions. This approach provides a more reliable and coordinate-independent way to identify regions prone to shocks, particularly when compression is significant overall but not dominant along any single axis. 

Additionally, using the divergence helps prevent misidentification of shock regions in cases where large negative directional derivatives
(e.g., $ \partial u / \partial x$) 
may be large and negative, but are nearly canceled by equally large positive components in other directions.
Such scenarios, common in contact discontinuities, could lead to false positives if only individual gradient terms are considered, whereas the divergence accurately indicates the absence of strong net compression. An example of this situation is illustrated in Problem 8 of the results section.

The proposed modifications of the Jacobian matrices for LLPINNs are
\begin{align}
     &\mathbf{\tilde{A}} = \mathbf{R} \tilde{\Lambda}\mathbf{R}^{-1}, \label{eq:Atilde2D}\\
     &\mathbf{\tilde{B}} = \mathbf{S} \tilde{M}\mathbf{S}^{-1}, \label{eq:Btilde2D}
\end{align}
where $\mathbf{R}$ and $\mathbf{S}$ are the matrices of right eigenvectors of $\mathbf{\tilde{A}}$ and $\mathbf{\tilde{B}}$, respectively, whereas the matrices $\tilde{\Lambda}$ and $\tilde{M}$ are
\begin{align}
    &\tilde{\Lambda} = \text{diag}\left(\tilde{\lambda}_-,\lambda_{01},\lambda_{02},\tilde{\lambda}_+ \right), \\
    &\tilde{M} = \text{diag}\left(\tilde{\mu}_-,\mu_{01},\mu_{02},\tilde{\mu}_+ \right),
\end{align}
with 
\begin{align}
    &\tilde{\lambda}_i=
    \begin{cases}
        s_i, \quad (x,y,t) \in \Omega_{sx} \land \lambda_i \left(\mathbf{U}_L \right) > s_i > \lambda_i \left(\mathbf{U}_R \right), \\
        \lambda_i, \quad \text{otherwise},
    \end{cases} \\
     &\tilde{\mu}_i=
    \begin{cases}
        s_i, \quad (x,y,t) \in \Omega_{sy} \land \mu_i \left(\mathbf{U}_L \right) > s_i > \mu_i \left(\mathbf{U}_R \right), \\
        \mu_i, \quad \text{otherwise},
    \end{cases}
\end{align}
being $s_i$ the shock speeds obtained for a 1D problem, defined in \eqref{eq:s1_euler} and \eqref{eq:s3_euler}. Note that the $L$ and $R$ subindices denote the \emph{down} and \emph{up} directions for the $\mu_i$ eigenvalues, respectively.

On the other hand, for the LRPINNs we calculate the Roe averages
\begin{align}
    \tilde{u} = \frac{\sqrt{\rho_L}u_L + \sqrt{\rho_R}u_R}{\sqrt{\rho_L} + \sqrt{\rho_R}}, \\
    \tilde{v} = \frac{\sqrt{\rho_L}v_L + \sqrt{\rho_R}v_R}{\sqrt{\rho_L} + \sqrt{\rho_R}}, \ \\
    \tilde{H} = \frac{\sqrt{\rho_L}H_L + \sqrt{\rho_R}H_R}{\sqrt{\rho_L} + \sqrt{\rho_R}},
\end{align}
and then we compute the modified Jacobian matrices
\begin{align}
    &\mathbf{\tilde{A}} =
    \begin{cases}
        \mathbf{A}\left(\tilde{u},\tilde{v},\tilde{H}\right), \quad (x,y,t) \in \Omega_{sx} \\
        \mathbf{A} \quad \text{otherwise},
    \end{cases}\\
    &\mathbf{\tilde{B}} =
    \begin{cases}
        \mathbf{B}\left(\tilde{u},\tilde{v},\tilde{H}\right), \quad (x,y,t) \in \Omega_{sy} \\
        \mathbf{B} \quad \text{otherwise}.
    \end{cases}
\end{align}
Here, $\mathbf{A}\left(\tilde{u},\tilde{v},\tilde{H}\right)$ and $\mathbf{B}\left(\tilde{u},\tilde{v},\tilde{H}\right)$ represent the matrices $\mathbf{A}$ and $\mathbf{B}$ defined in \eqref{eq:A2D} and \eqref{eq:B2D}, evaluated at the Roe averages. Note again that the subindices $L$ and $R$ denote the \emph{down} and \emph{up} directions for the matrix $\mathbf{B}$, respectively.

The Algorithms \ref{alg:LLPINNs2D} and \ref{alg:lrpinns2D} provide a detailed description of the steps involved in our proposed extension of the LLPINNs and LRPINNs to the 2D Euler equations.

\begin{figure}
\begin{algorithm}[H]
\caption{Generalized LLPINNs for 2D Euler equations}\label{alg:LLPINNs2D}
\begin{algorithmic}
   \State \textbf{Input:} $\Omega^{(D)}$, $\Omega^{(I)}$, $\Omega^{(B)}$, $M_x$, $M_y$, $\Delta x$, $\Delta y$, $\omega \equiv(\omega_D,\omega_I,\omega_B)$, $\eta$.
  \State \textbf{Output:} PINN solution $\mathbf{U}\left(x,t; \Theta\right)$
  \State Initialize trainable variables $\Theta$
  \While{$\mathcal{J}(\Theta) > \eta$}
  \State Calculate $\partial_x u_ \equiv \frac{\partial u (\mathbf{x})}{\partial x}$ and $\partial_y v_ \equiv \frac{\partial v (\mathbf{x})}{\partial y}$ for each point $\mathbf{x} \in \Omega^{(D)}$
  \State Determine the points $\mathbf{x}_s^{(x)} = \left(x_s,y_s, t_s \right) \in \Omega_{sx}$ and $\mathbf{x}_s^{(y)} = \left(x_s,y_s, t_s \right) \in \Omega_{sy}$ using the criteria \eqref{eq:xshock_criterion} and \eqref{eq:yshock_criterion}.
  \State Compute $\mathbf{x}_{L}^{(x)} = (x_s-\Delta x, y_s, t_s)$, $\mathbf{x}_{R}^{(x)} = (x_s+\Delta x,y_s,t_s)$, $\mathbf{x}_{L}^{(y)} = (x_s, y_s - \Delta y, t_s)$ and $\mathbf{x}_{R}^{(y)} = (x_s,y_s + \Delta y,t_s)$
  \State Obtain left and right states for each coordinate direction: $\mathbf{U}_L^{(x)} = \mathbf{U}\left(\mathbf{x}_L^{(x)};\Theta \right)$, $\mathbf{U}_R^{(x)} = \mathbf{U}\left(\mathbf{x}_R;\Theta \right)$, $\mathbf{U}_L^{(y)} = \mathbf{U}\left(\mathbf{x}_L^{(y)};\Theta \right)$, $\mathbf{U}_R^{(y)} = \mathbf{U}\left(\mathbf{y}_R;\Theta \right)$.
  \State Compute the possible shock velocities $s_-$ and $s_+$, given by the expressions \eqref{eq:s1_euler} and \eqref{eq:s3_euler}, for each coordinate direction.
  \State Use Lax entropy conditions \eqref{eq:lax_entropy} for each coordinate direction to determine if $\textbf{x}_s^{(x)}$ belongs to the $\lambda_-$ or to the $\lambda_+$ shock, and to determine if $\textbf{x}_s^{(y)}$ belongs to the $\mu_-$ or to the $\mu_+$ shock
  \State Change $\lambda_-$ or $\lambda_+$, and $\mu_-$ or $\mu_+$, by $s_-$ and $s_+$ in each coordinate direction, according to the result obtained in the previous step.
  \State Compute the modified Jacobians $\tilde{\mathbf{A}}$ and $\tilde{\mathbf{B}}$ given by \eqref{eq:Atilde2D} and \eqref{eq:Btilde2D}.
  \State Calculate $\mathcal{L}_D$, defined in \eqref{eq:Ld}, with the PDE residuals $\mathcal{D}\left[\mathbf{U} \right](\mathbf{x},t) = \frac{\partial \mathbf{U}}{\partial t} + \mathbf{\tilde{A}}\left(\mathbf{U} \right)\frac{\partial \mathbf{U}}{\partial x} + + \mathbf{\tilde{B}}\left(\mathbf{U} \right)\frac{\partial \mathbf{U}}{\partial y}$ for each point $(\mathbf{x},t) \in \Omega^{(D)}$.
  \State Calculate $\mathcal{L}_I$ and $\mathcal{L}_B$, respectively defined in \eqref{eq:Li} and \eqref{eq:Lb}. for the points $(\mathbf{x},t)$ belonging to $\Omega^{(I)}$ and $\Omega^{(R)}$.
  \State Calculate $\mathcal{J}(\Theta)$ using \eqref{eq:loss} and perform an optimization step.
  \EndWhile
\end{algorithmic}
\end{algorithm}
\end{figure}

\begin{figure}
\begin{algorithm}[H]
\caption{LRPINNs for 2D Euler equations}\label{alg:lrpinns2D}
\begin{algorithmic}[1]
  \State \textbf{Input:} $\Omega^{(D)}$, $\Omega^{(I)}$, $\Omega^{(B)}$, $M_x$, $M_y$, $\Delta x$, $\Delta y$, $\omega \equiv(\omega_D,\omega_I,\omega_B)$, $\eta$.
  \State \textbf{Output:} PINN solution $\mathbf{U}\left(x,t; \Theta\right)$
  \State Initialize trainable variables $\Theta$
  \While{$\mathcal{J}(\Theta) > \eta$}
  \State Calculate $\partial_x u_ \equiv \frac{\partial u (\mathbf{x})}{\partial x}$ and $\partial_y v_ \equiv \frac{\partial v (\mathbf{x})}{\partial y}$ for each point $\mathbf{x} \in \Omega^{(D)}$
  \State Determine the points $\mathbf{x}_s^{(x)} = \left(x_s,y_s, t_s \right) \in \Omega_{sx}$ and $\mathbf{x}_s^{(y)} = \left(x_s,y_s, t_s \right) \in \Omega_{sy}$ using the criteria \eqref{eq:xshock_criterion} and \eqref{eq:yshock_criterion}.
  \State Compute $\mathbf{x}_{L}^{(x)} = (x_s-\Delta x, y_s, t_s)$, $\mathbf{x}_{R}^{(x)} = (x_s+\Delta x,y_s,t_s)$, $\mathbf{x}_{L}^{(y)} = (x_s, y_s - \Delta y, t_s)$ and $\mathbf{x}_{R}^{(y)} = (x_s,y_s + \Delta y,t_s)$
  \State Obtain left and right states for each coordinate direction: $\mathbf{U}_L^{(x)} = \mathbf{U}\left(\mathbf{x}_L^{(x)};\Theta \right)$, $\mathbf{U}_R^{(x)} = \mathbf{U}\left(\mathbf{x}_R;\Theta \right)$, $\mathbf{U}_L^{(y)} = \mathbf{U}\left(\mathbf{x}_L^{(y)};\Theta \right)$, $\mathbf{U}_R^{(y)} = \mathbf{U}\left(\mathbf{y}_R;\Theta \right)$.
  \State Calculate the corresponding Roe averages in each coordinate direction
  \State Compute the modified Jacobian $\tilde{\mathbf{A}}$ and $\tilde{\mathbf{B}}$ in terms of the Roe averages given at the previous step.
  \State Calculate $\mathcal{L}_D$, defined in \eqref{eq:Ld}, with the PDE residuals $\mathcal{D}\left[\mathbf{U} \right](\mathbf{x},t) = \frac{\partial \mathbf{U}}{\partial t} + \mathbf{\tilde{A}}\left(\mathbf{U} \right)\frac{\partial \mathbf{U}}{\partial x} + + \mathbf{\tilde{B}}\left(\mathbf{U} \right)\frac{\partial \mathbf{U}}{\partial y}$ for each point $(\mathbf{x},t) \in \Omega^{(D)}$.
  \State Calculate $\mathcal{L}_I$ and $\mathcal{L}_B$, respectively defined in \eqref{eq:Li} and \eqref{eq:Lb}. for the points $(\mathbf{x},t)$ belonging to $\Omega^{(I)}$ and $\Omega^{(R)}$.
  \State Calculate $\mathcal{J}(\Theta)$ using \eqref{eq:loss} and perform an optimization step.
  \EndWhile
\end{algorithmic}
\end{algorithm}
\end{figure}

\section{Results}

In the subsequent sections, we provide results for various benchmark problems related to the physical applications previously discussed. Specifically, we evaluate the performance of three algorithms: a standard method that uses the Jacobian
 ($\mathbf{\tilde{A}} = \mathbf{A}$) across the entire domain, our generalized version of LLPINNs as outlined in Algorithms 1 and 3, and the approach based on approximate Riemann solvers (specifically Roe), referred to as LRPINNs and described in Algorithms 2 and 4. 

Since neural networks are continuous and differentiable, they exhibit better convergence to continuous functions. To accelerate convergence, we initially train (1500–2000 iterations in the 1D problems, and 3000 iterations in the 2D problems) the PINN using a viscous version of the problem 
\begin{equation}
    \frac{\partial \mathbf{U}}{\partial t} + \frac{\partial \mathbf{F}(\mathbf{U})}{\partial x} + \frac{\partial \mathbf{G}(\mathbf{U})}{\partial y} = \nu \nabla^2 \mathbf{U},
\end{equation}
where $\nabla^2 \mathbf{U}$ denotes the Laplacian vector of $\mathbf{U}$ and
$\nu$ is a small viscosity of the order of $10^{-3}$.
The neural network's prediction from this step is then used to re-initialize the network for the subsequent training phase with the non-viscous ($\nu = 0$) equation. 
Note that these are not the physical Navier-Stokes equations but a modified Euler equation with viscous terms applied to all variables. 

\begin{table*}[!htbp]
    \begin{tabular}{c c c c c c c c c c c c}
    
         Problem & Dimension & Architecture & $a$ & $M$ & $M_x$ & $\Delta x$ & $M_y$ & $\Delta y$
         & $\nu/10^{-3}$ & \makecell{Viscous \\ iters.} \\ \hline
         1 & 1D & 7L20N & $0.05$ & $0.001$ & $-$ & $0.02$ & $-$ & $-$ & $3$ & $1500$ \\
         2 & 1D & 7L20N & $0.05$ & $0.001$ & $-$ & $0.02$ & $-$ & $-$ & $3$ & $1500$\\
         3 & 1D & 7L20N & $0.1$ & $0.001$ & $-$ & $0.015$ & $-$ & $-$ & $3$ & $2000$ \\
         4 & 1D & 6L20N & $0.05$ & $0.005$ & $-$ & $0.02$ & $-$ & $-$ & $0.5$ & $2000$ \\
         $5$ & 1D & 6L20N & $0.05$ & $0.001$ & $-$ & $0.008$ & $-$ & $-$ & $0.3$ & $1500$ \\
         6 & 2D & 7L20N & $0.1$ & $2 \times 10^{-5}$ & $10^{-5}$ & $0.025$ & $10^{-5}$ & $0.025$ & $2$ & $3000$  \\
         7 & 2D & 6L30N & $0.05$ & $0.02$ & $0.02$ & $0.01$ & $0.02$ & $0.01$ & $2$ & $3000$ \\ 
         8 & 2D & 6L30N & $0.05$ & $0.7$ & $0.7$ & $0.01$ & $0.7$ & $0.01$ & $1$ & $3000$
    \end{tabular}
    \caption{Hyperparameters considered for each problem. Here, $m$L$n$N denotes a layer with $m$ hidden layers and $n$ neurons at each layer. }
    \label{tab:hyperparameters}
\end{table*}

\subsection{1D problems}
We start with the one-dimensional problems. In this case, the LLPINN and LRPINN methodologies correspond to Algorithms \ref{alg:LLPINNs} and \ref{alg:lrpinns}, respectively. To improve convergence, we multiply the PDE residuals by the gradient-annihilated factor introduced in \cite{ferrer2024gradient}
\begin{equation}
    \alpha = \frac{1}{1 + a\left|\frac{\partial u}{\partial x} \right|},
\end{equation}
with $a$ being a user-defined constant, to homogenize the PDE residuals in the entire domain. 
Table \ref{tab:hyperparameters} shows all the hyperparameters considered for each of the problems listed below.
The loss function is given by a \eqref{eq:loss} with $\omega_R=1$ and $\omega_I = 10$.

The training set consists of points $\left( x,t \right) \in \Omega_R \times \Omega_I$, with $\Omega_R$ and $\Omega_I$ defined in \eqref{set:OmegaD} and \eqref{set:omegaI}. These points are randomly chosen following a uniform distribution, and changed every 500 iterations at most\footnote{We say ``at most'' because the optimization process for a given set of points may be interrupted before completing $500$ iterations if the optimizer does not find a proper value of step-length $\alpha_k$. When this happens, we change the training set and continue the training process.}.

\subsubsection{Euler equations}

\paragraph{Problem 1. \textbf{Double shock wave}} 
We consider the double shock wave problem of \cite{LLPINNs} (\emph{Case 5}) with the initial conditions
\begin{equation*}
    (\rho,u,p)=
    \begin{cases}
        &(1.0,0.5,1.0), \quad  x < 0, \\
        &(0.5,0.0,1.0), \quad x>0.
    \end{cases}
\end{equation*}

\begin{figure*}
    \begin{subfigure}[t]{0.93\columnwidth}
        \centering
        \caption{}
        \includegraphics[width=\columnwidth]{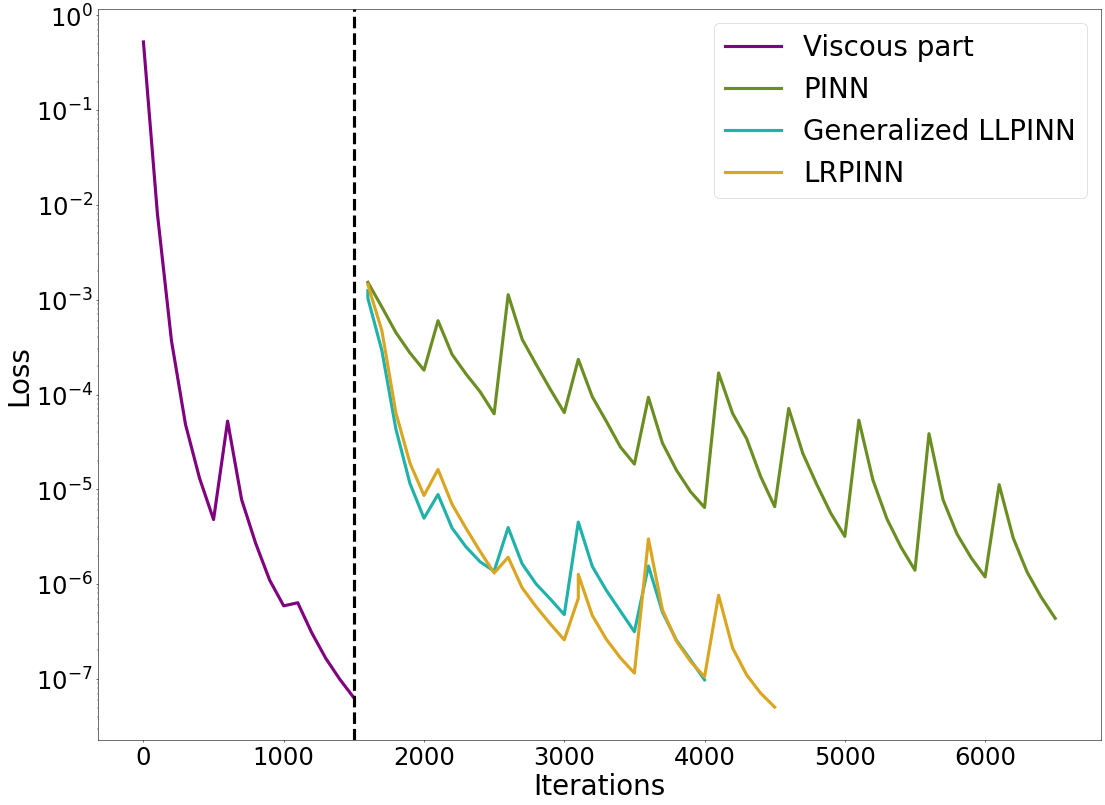}
    \end{subfigure}
    \begin{subfigure}[t]{0.93\columnwidth}
        \centering
        \caption{}
        \includegraphics[width=\columnwidth]{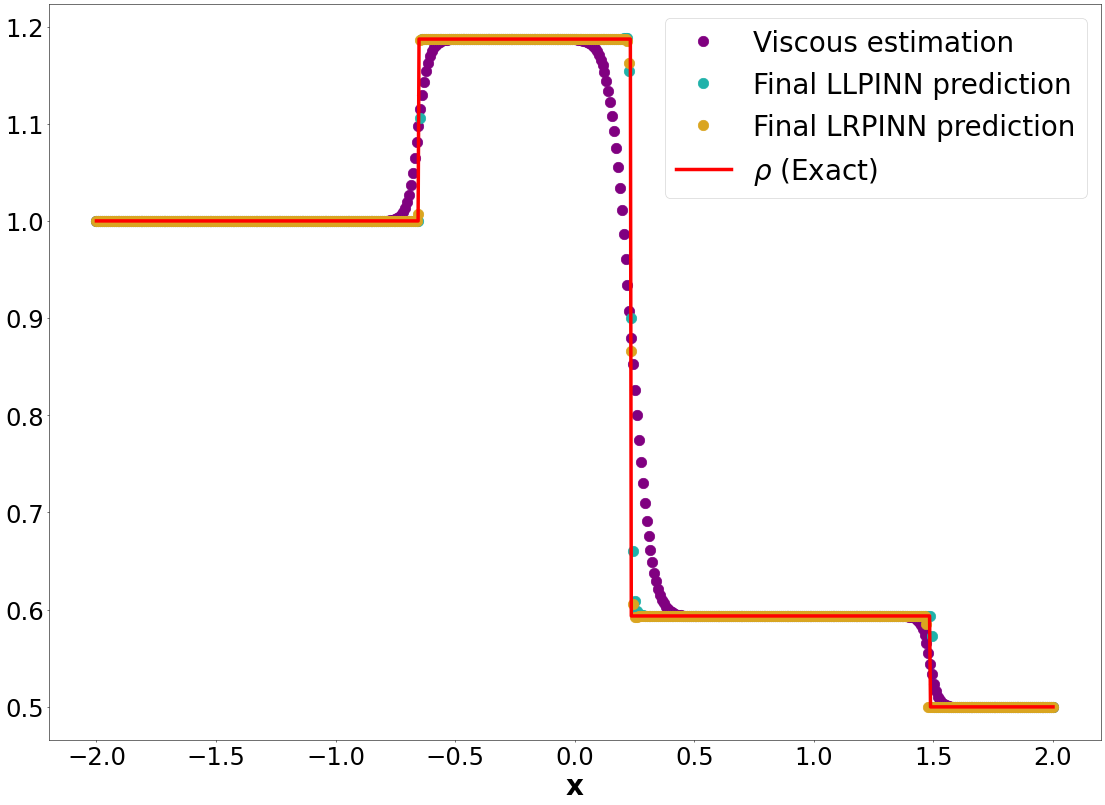}
    \end{subfigure}
    \caption{(\textbf{a}) Loss function evolutions with the iterations for the different methodologies proposed here: the \emph{classical} PINN (labeled here as PINN), our generalized version of LLPINNs (Generalized LLPINN), and LRPINNs. The viscous part of the training process is also plotted in purple and is common for all three methodologies. (\textbf{b}) Viscous estimation of the density, which corresponds to the prediction of the different models at the end of the viscous part of the training process, and the final predictions obtained with the LLPINN and the LRPINN.}
    \label{fig:loss_2_shocks_euler}
\end{figure*}

\begin{figure*}
    \centering
    \includegraphics[width=0.93\textwidth]{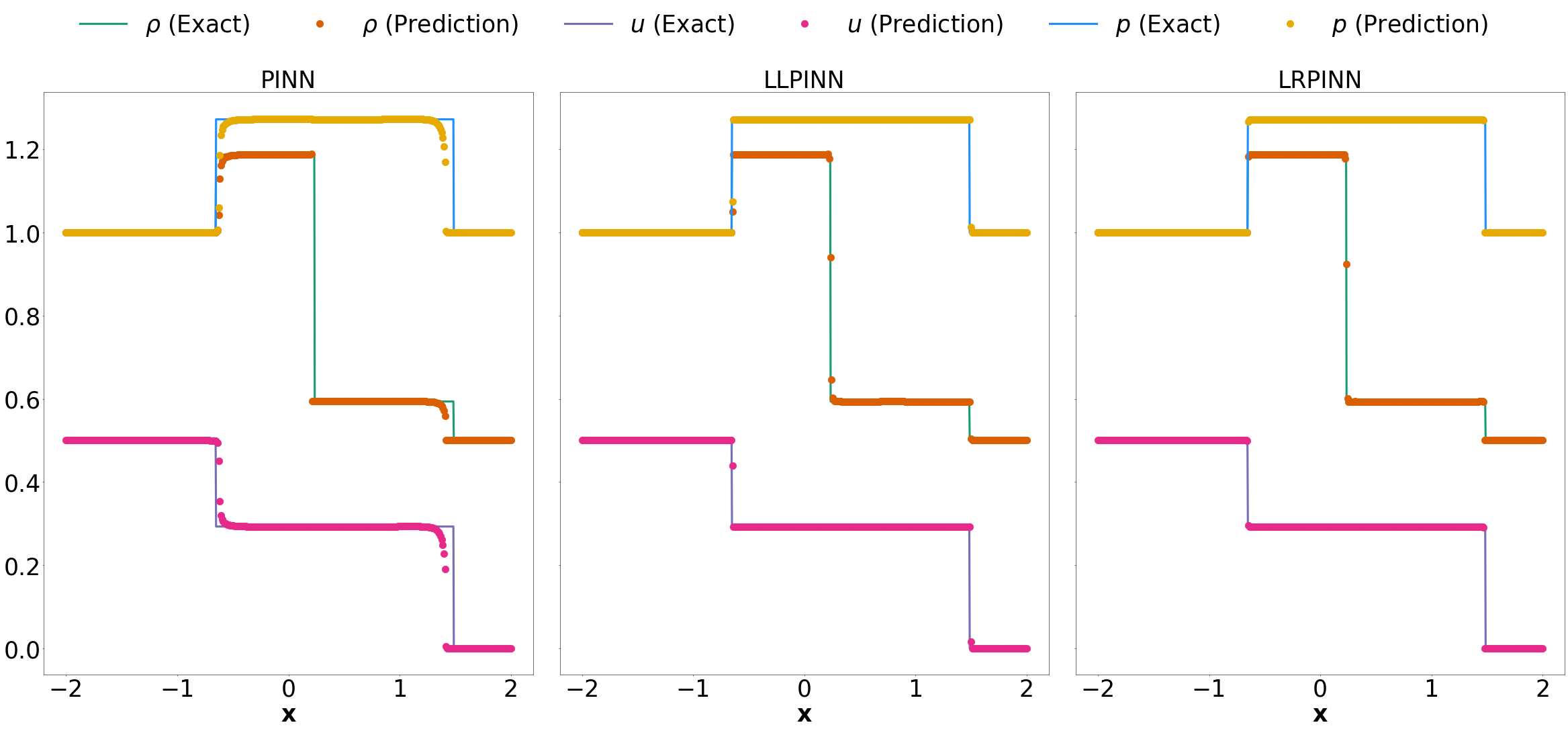}
    \caption{Final predictions for each algorithm (PINN, LLPINN and LRPINN) for the double shock problem (Problem 1). The neural network is evaluated at $400$ spatial points at the final time. The exact solution is also plotted for reference.}\label{fig:double_shock_euler_predictions}
\end{figure*}

\begin{figure*}
    \begin{subfigure}[t]{0.93\columnwidth}
        \centering
        \caption{}
        \includegraphics[width=\columnwidth]{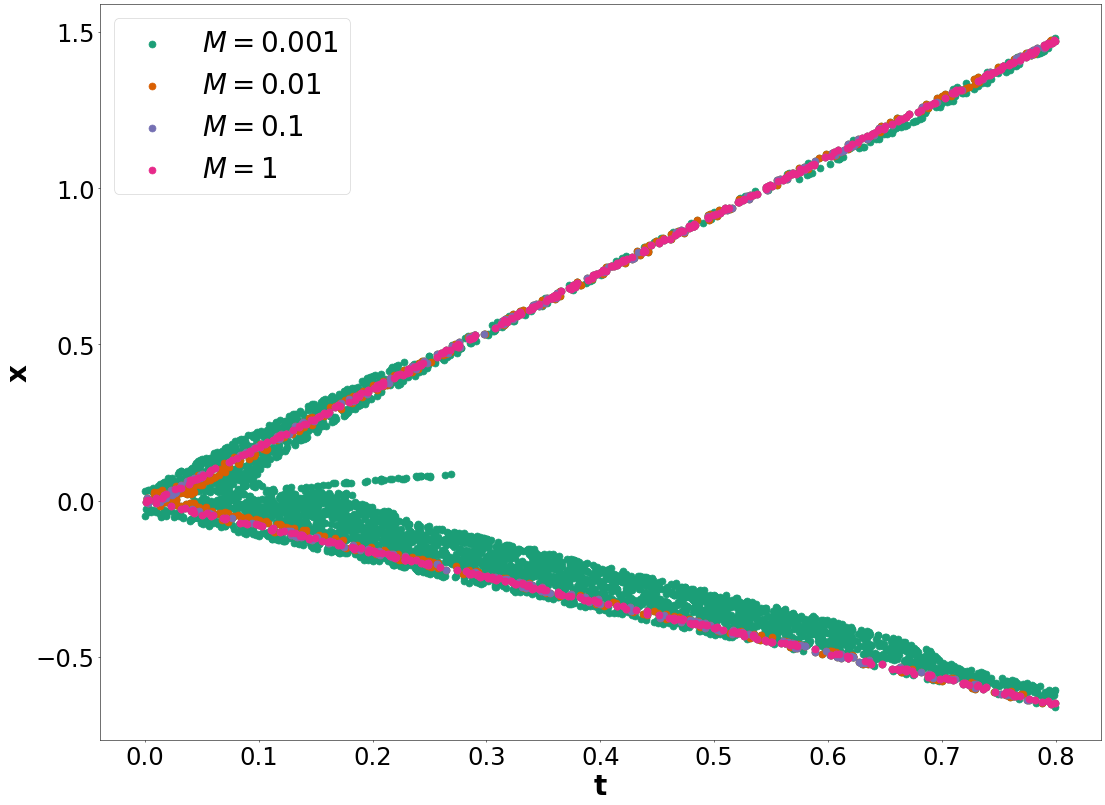}
    \end{subfigure}
    \begin{subfigure}[t]{0.93\columnwidth}
        \centering
        \caption{}
        \includegraphics[width=\columnwidth]{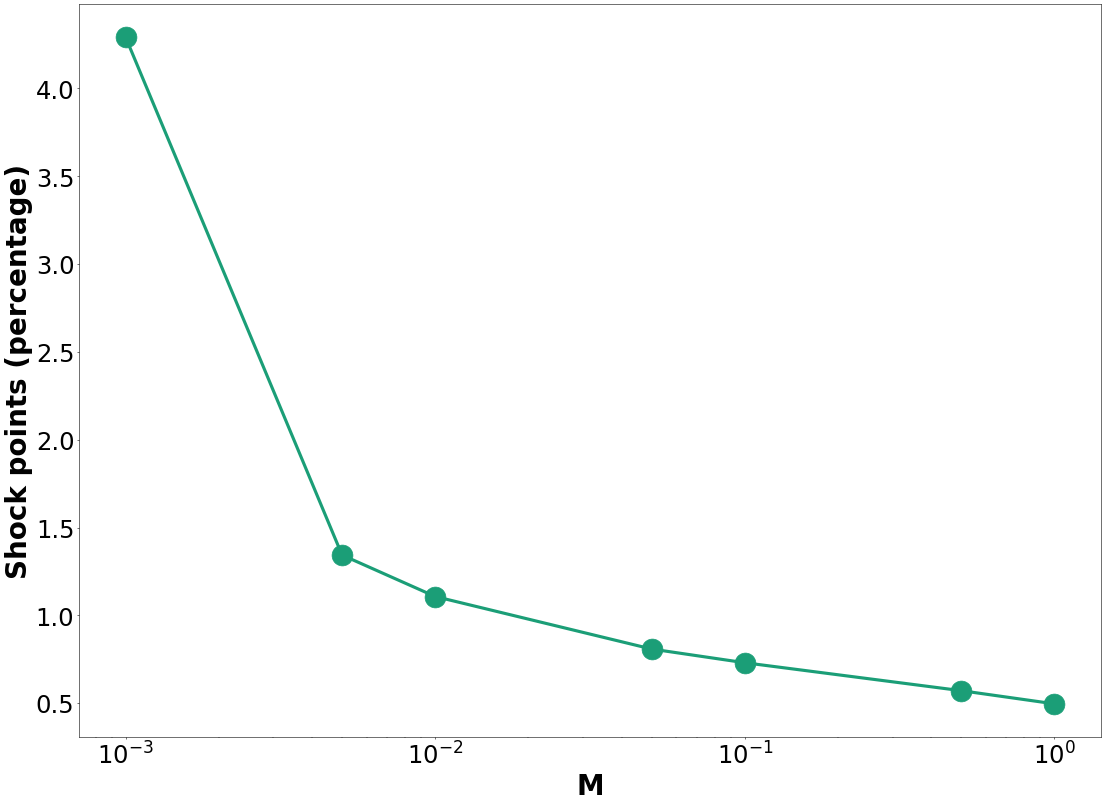}
    \end{subfigure}
    \caption{(\textbf{a}) Detected shock points for different values of $M$ from a given test set of random $10^5$ points. (\textbf{b}) Percentage of detected shock points with respect the total test set.}
    \label{fig:shock_points_1D}
\end{figure*}

In this problem, the solution corresponds to two shock waves moving in opposite directions from the initial discontinuity, located at $x=0$. The solution domain in this case is $\left(x,t\right) \in \left[-2,2 \right] \times \left[0,0.8 \right]$.

The left panel of Figure \ref{fig:loss_2_shocks_euler} shows the evolution of the loss function for the three methods. As anticipated, the latter 
two methods reduce the loss function much faster than standard PINNs, which do not account for the possibility of discontinuities. Including information about the solution of the Riemann problem (shock speed or jump conditions) in the Jacobian matrix enhances convergence speed and yields significantly improved results. To illustrate the effect of the initial pretraining, the right panel compares the density profiles at $t=0.8$ after the first 1500 iterations (viscous estimation) with the density profiles obtained at the end of the training process. Note that the network generates solutions for all times within $\left[0,0.8 \right]$.  Both shock waves, and particularly the contact wave, appear significantly sharper after completing the inviscid part of the training process.
Figure \ref{fig:double_shock_euler_predictions} compares the predictions of the three PINN versions with the exact solution for all variables. Both LLPINNs and LRPINNs accurately localize the two shock waves with sharp transitions, closely matching the exact solution. In contrast, the standard PINN yields a smoother solution profile and incorrectly locates the shocks, despite some similarity.

Finally, note from Table \ref{tab:hyperparameters} that we have selected a small but positive value for $M$. As discussed in the previous section, maintaining a low value of $M$ is preferable, as it ensures a sufficiently large set of points where Jacobian corrections are applied, while also allowing the optimization algorithm to make reasonable progress. The left panel of Figure \ref{fig:shock_points_1D} displays the shock points identified by the fully trained network, for a test set of $10^5$ points distributed across the domain, and for various values of $M$. The right panel shows the corresponding percentage of detected shock points relative to the total set. Both shock waves propagating in opposite directions are clearly visible. The number of detected shock points decreases as $M$ increases, ranging from $\sim 4\%$ for $M=0.001$
to less than $0.5\%$ for $M=1$. 

\begin{table}[h!]
    \centering
    \begin{tabular}{cccccc}
         Method & Points & $L^2_\rho$ & $L^2_u$ & $L^2_p$ & Time (s) \\ \hline 
         WENO-Z  & $100$ & $0.034$ & $0.045$ & $0.015$ & $2$ \\
         WENO-Z  & $200$ & $0.027$ & $0.035$ & $0.012$ & $8$ \\
         WENO-Z  & $400$ & $0.021$ & $0.030$ & $0.009$  & $35$      \\
         WENO-Z  & $800$ & $0.014$ & $0.019$ & $0.006$ & $140$      \\
         WENO-Z  &$1600$ & $0.011$ & $0.015$ & $0.005$ & $580$       \\
         LRPINN & - &  $0.011$ & $0.014$ & $0.005$ & $750$
    \end{tabular}
    \caption{Relative $L^2$ error norms of the errors (differences with the analytical solution) for the density, velocity, and pressure predictions obtained with a WENO-Z scheme\cite{WenoZ_github} for different resolutions and with the LRPINN. All the WENO-Z experiments have been done with a CFL number equal to 0.4. The LRPINN has been trained for $2500$ iterations in total, with $1000$ viscous iterations and another $1500$ for the non-viscous training phase. For reference, the last column shows the total execution time on the same hardware (a normal laptop).}
    \label{tab:errors}
\end{table}

To assess the accuracy of our numerical results, Table~\ref{tab:errors} reports the relative $L^2$ error norms for density, velocity, and pressure,
comparing the LRPINN predictions with those 
using a WENO-Z solver \cite{WENOZ} across different spatial grid resolutions.
Remarkably, the LRPINN demonstrates superior accuracy compared to the WENO-Z method up to a resolution of roughly 1000 spatial points.

\paragraph{\textbf{Problem 2. Sod shock tube}} The next problem we consider is the Sod shock tube problem (\emph{Case 4} of \cite{LLPINNs}), with the initial conditions
\begin{equation*}
    (\rho,u,p)=
    \begin{cases}
        &(1.0,0.0,1.0), \quad x < 0, \\
        &(0.125,0.0,0.1), \quad x>0.
    \end{cases}
\end{equation*}

\begin{figure*}
    \centering
    \includegraphics[width=0.93\textwidth]{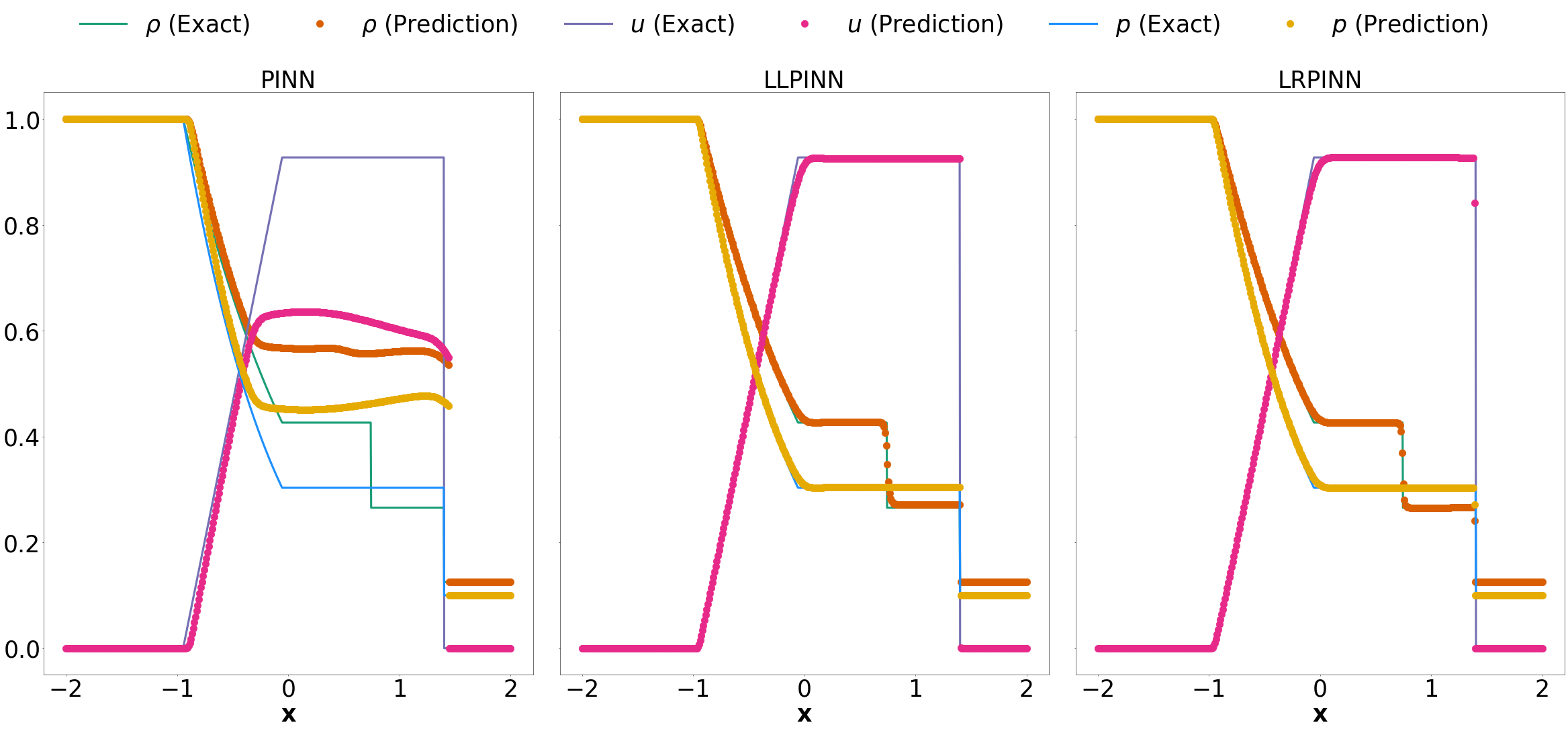}
    \caption{Same as Figure \ref{fig:double_shock_euler_predictions} for the Sod shock tube problem.}\label{fig:sod_euler_predictions}
\end{figure*}

\begin{figure*}
    \begin{subfigure}[t]{.93\columnwidth}
        \centering
        \caption{}
        \includegraphics[width=\columnwidth]{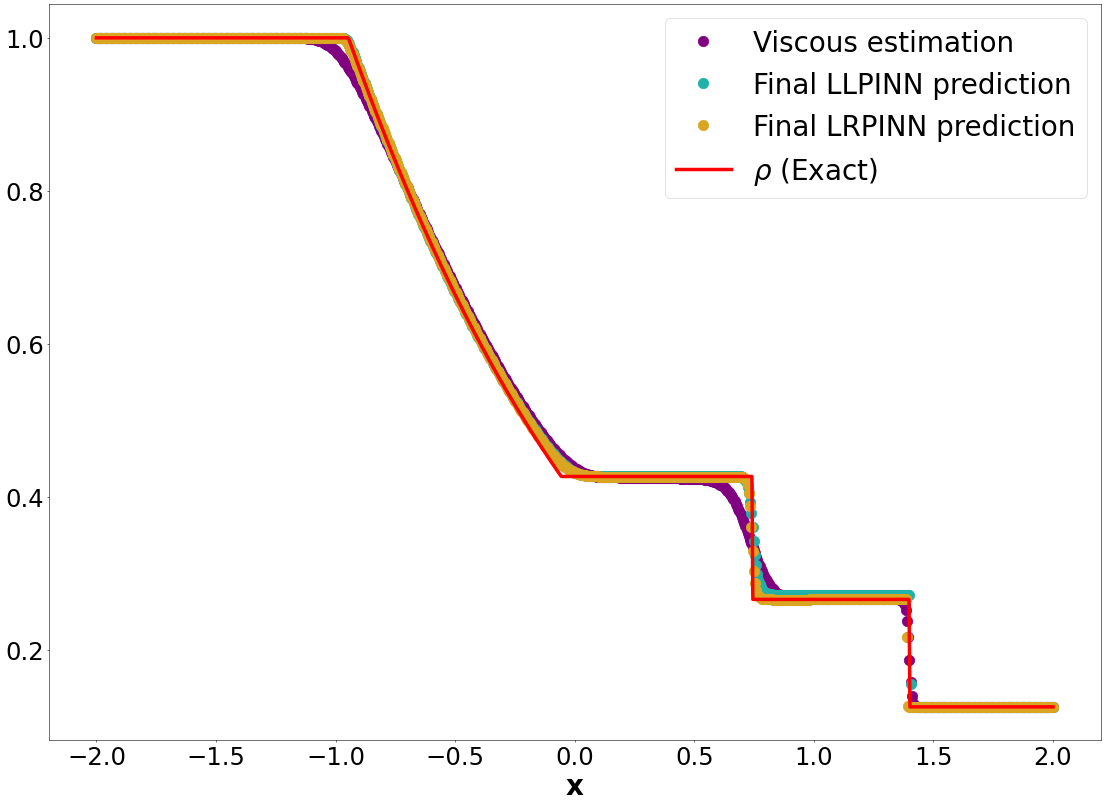}
    \end{subfigure}
    \begin{subfigure}[t]{.93\columnwidth}
        \centering
        \caption{}
        \includegraphics[width=\columnwidth]{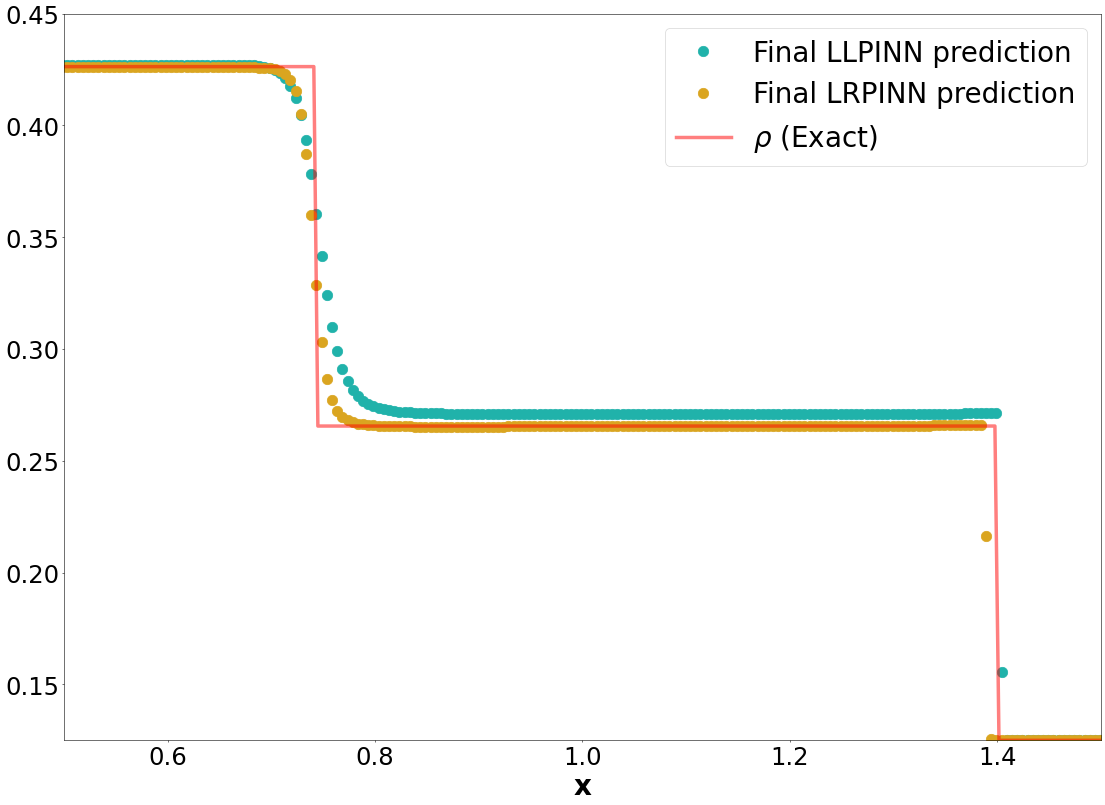}
    \end{subfigure}
    \caption{(\textbf{a}) Same as Figure \ref{fig:loss_2_shocks_euler} (\textbf{b}) for the Sod shock tube. The exact solution is also plotted for reference. (\textbf{b}) A zoom at the region between the contact and shock waves, with the LLPINN and LRPINN predictions of the density, and the exact solution.}
    \label{fig:density_comparison_LLPINN_lrpinn_euler_sod}
\end{figure*}

The solution consists of a shock wave moving to the right and a rarefaction wave to the left of the initial discontinuity. The domain of the PINN prediction is again $\left(x,t\right) \in \left[-2,2 \right] \times \left[0,0.8 \right]$.

Figure \ref{fig:sod_euler_predictions} compares the final predictions of the PINN, LLPINN, and LRPINN methodologies with the exact solution of this problem. We can see that, although the PINN detects the presence of the shock wave, its location is close but not accurate, and the jump conditions are notoriously violated. Significant discrepancies also appear in the rarefaction wave. Most notably, and concerningly, the contact wave in the density has been entirely lost. This result highlights an important insight: even though the modifications to the Jacobian matrix are applied only in regions of potential shock formation, they have a global effect on convergence.

This significant issue with the PINN is not an isolated incident but occurs consistently. All three networks are pretrained identically, as depicted in the left panel of 
Figure \ref{fig:density_comparison_LLPINN_lrpinn_euler_sod}, which compares the viscous estimation (obtained after $1500$ iterations with $\nu = 0.003$) with the final LLPINN and LRPINN predictions. 
Although the viscous solution offers a reasonable starting point for training, the standard PINN approach fails to capture discontinuous solutions when viscosity is set to zero. Consequently, hereafter we omit results obtained with the standard PINN, as it frequently fails to converge satisfactorily.

In contrast, LLPINNs and LRPINNs exhibit excellent performance, with good resolution of shock and contact waves. However, as highlighted in the right panel of 
Figure \ref{fig:density_comparison_LLPINN_lrpinn_euler_sod}, which zooms in on the density profile, a subtle discrepancy is observed in the \emph{post}-shock density state, specifically between the contact and shock waves, where the density value is slightly higher than the correct value.
This issue, also noted but not discussed in detail in \cite{LLPINNs} 
(see their Figure 14) is evident in this case. As shown in the next example, this discrepancy results in a failure to conserve mass, leading to subsequent losses in momentum and energy conservation during the evolution.

\paragraph{\textbf{Problem 3: Lax shock tube}}

The next problem we consider for the 1D Euler equations is the Lax shock tube, whose initial condition is
\begin{equation*}
    (\rho,u,p)=
    \begin{cases}
        &(0.445,0.698,3.528), \quad x < 0, \\
        &(0.5,0.0,0.571), \quad x>0.
    \end{cases}
\end{equation*}
This initial condition yields a solution resembling the previous case but with a stronger shock that propagates at a higher speed. The domain for the PINN prediction is $\left(x, t\right) \in \left[-2.5, 2.5\right] \times \left[0, 0.8\right]$. The viscous phase of the training process, with $\nu = 0.004$, lasts for $2000$ iterations. 

\begin{figure*}
    \centering
    \includegraphics[width=0.93\textwidth]{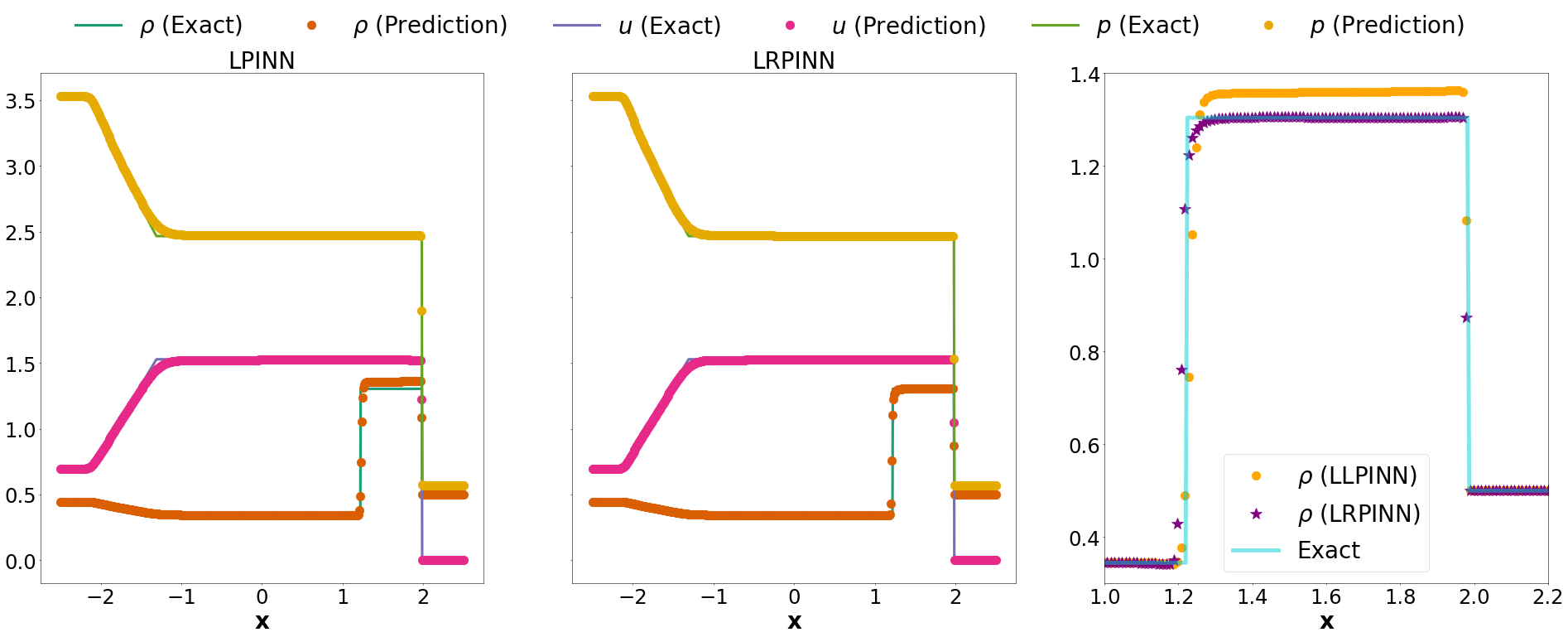}
    \caption{LLPINN and LRPINN final predictions for all the variables (left and central panels, respectively), with a zoom in the predicted density between the contact and the shock waves at the right panel. The plots are generated by taking $500$ points at the final time.}\label{fig:lax_euler_predictions}
\end{figure*}

Figure \ref{fig:lax_euler_predictions} presents the results obtained with both LLPINNs and LRPINNs. In this test case, a more pronounced deviation is observed in the LLPINN’s density prediction within the region between the contact and shock waves, whereas the LRPINN accurately captures the discontinuities in all variables, including the density. The right panel of the figure provides a zoomed-in view of the density profile in this region, clearly highlighting the difference between the two approaches.

\begin{figure}
    \centering
    \includegraphics[width=\columnwidth]{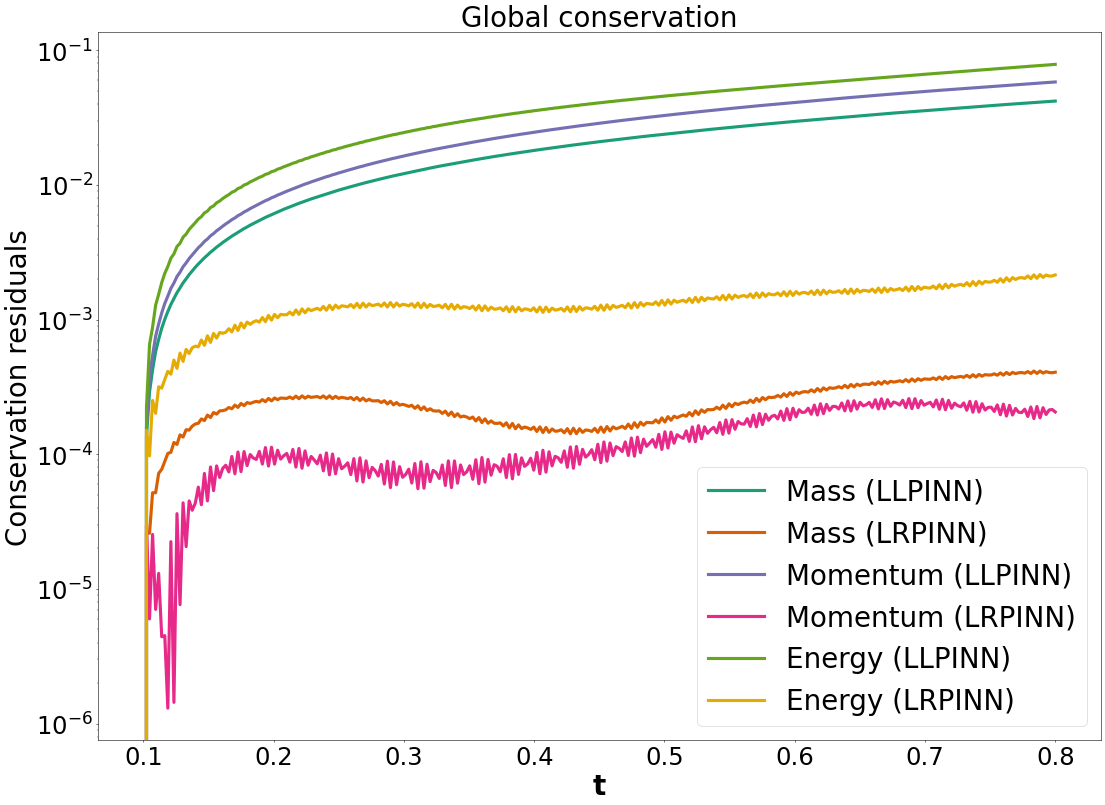}
    \caption{Difference between the conserved quantities and their initial values as a function of time obtained with the LLPINN and LRPINN methods for the Lax shock tube problem.}
    \label{fig:mass_residuals_lax}
\end{figure}

This discrepancy is closely linked to the fact that the Jacobian modification in LLPINNs does not guarantee the enforcement of conservation laws across discontinuities. To illustrate this, Figure \ref{fig:mass_residuals_lax} shows the difference between the total integrated conserved quantities in the domain and its initial value as a function of time. While this residual remains nearly constant and on the order of $10^{-3}$ when using the LRPINN, it increases over time in the case of the LLPINN.

\paragraph{\textbf{Problem 4: Sod shock tube in a moving reference frame}}

To end with the 1D Euler equations, we consider the initial condition
\begin{equation*}
    (\rho,u,p)=
    \begin{cases}
        &(3.0,0.9,3.0), \quad x < 0, \\
        &(1.0,0.9,1.0), \quad x>0,
    \end{cases}
\end{equation*}
as proposed by LeVeque \cite{leveque1998computational}. The domain for the PINN prediction is $\left(x, t\right) \in \left[-2.5, 2.5\right] \times \left[0, 0.8\right]$. The viscous phase of the training process, with $\nu = 0.004$, lasts for $2000$ iterations. 

In the same way as the Sod shock tube problem, the solution from left to right consists of a rarefaction wave, followed by a contact discontinuity, and a shock wave. However, the main difference now is that the rarefaction wave becomes \emph{transonic}, that is, the eigenvalue $\lambda_1 = u-a$ changes sign along the rarefaction. The traditional Roe solver fails to capture the correct physical solution in the presence of transonic rarefaction waves, requiring modifications to the numerical flux to ensure entropy consistency in such cases. Classical examples of these modifications are the Harten\cite{HARTEN1983357}, the Harten-Hyman\cite{HARTEN1983235} and the LeVeque\cite{LeVeque_2002} formulations.

\begin{figure*}
    \centering
    \includegraphics[width=0.93\textwidth]{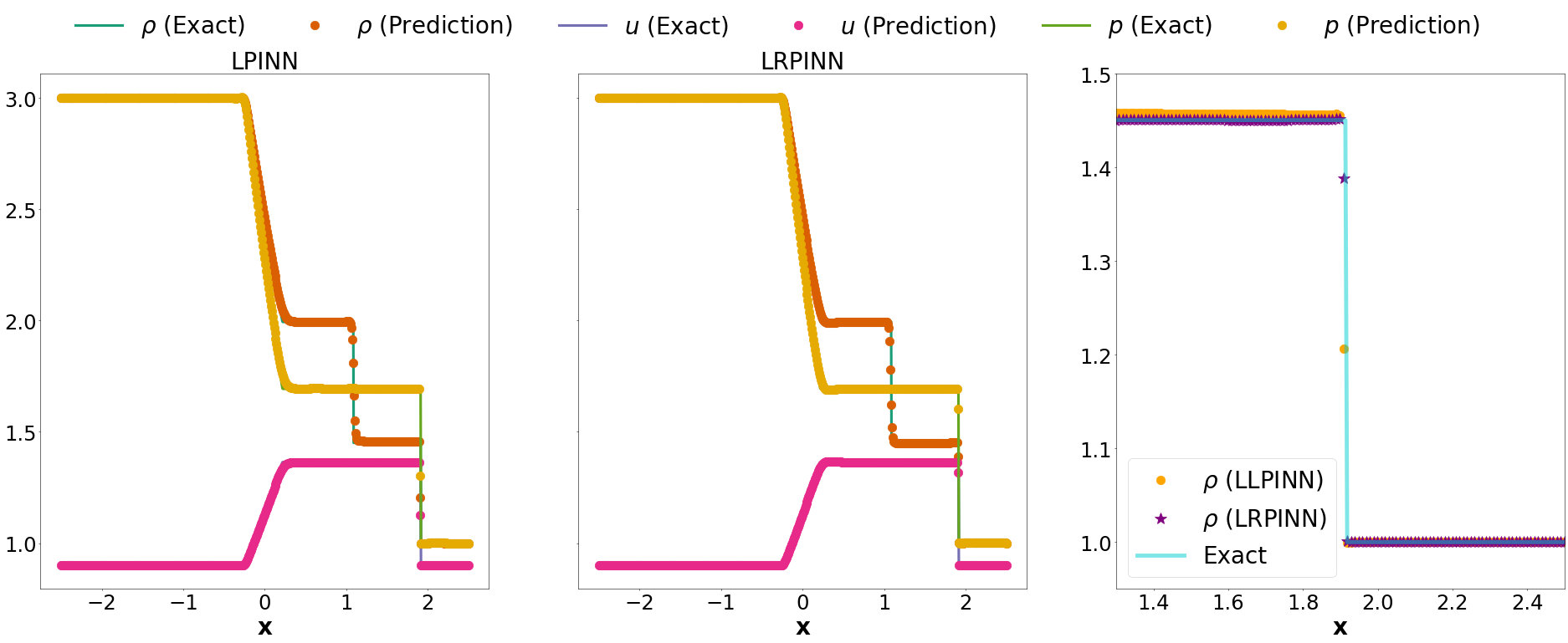}
    \caption{Same as Figure \ref{fig:lax_euler_predictions} for the Sod shock tube problem in a moving reference frame.}\label{fig:solution_moving_sod}
\end{figure*}

Since LRPINNs incorporate the Roe-type Jacobian modification, it is natural to investigate whether they also struggle to accurately capture solutions involving transonic rarefaction waves. However, as shown in the central panel of Figure~\ref{fig:solution_moving_sod}, the LRPINN successfully predicts the correct solution. This is because the original Jacobian matrix is \emph{not} replaced by the Roe Jacobian in regions with rarefaction waves, where $\frac{\partial u}{\partial x} \geq 0$, but only in regions indicative of potential shock formation, characterized by negative values of $\frac{\partial u}{\partial x}$. The left panel of Figure~\ref{fig:solution_moving_sod} shows the solution obtained with the generalized LLPINN, whereas the right panel shows a zoom of the predicted densities in the region between the contact and the shock waves obtained with both methods. Again, a small but noticeable discrepancy appears in the density prediction for the LLPINN.

\subsubsection{Special relativity hydrodynamics}

\paragraph{\textbf{Problem 4: Relativistic Sod shock tube}}
We now shift our focus to the more challenging equations of special relativistic hydrodynamics, using the same initial conditions as in Problem 1 from \cite{ferrer2024gradient}:
\begin{equation*}
    (\rho,u,p)=
    \begin{cases}
        &(1.0,0.5,1.0), \quad x < 0.5, \\
        &(0.125,0.5,0.1), \quad x>0.5
    \end{cases}
\end{equation*}
which are the same as the non-relativistic Sod shock tube for the density and pressure, but now the flow velocity is nonzero. The solution for $\gamma=5/3$ has a rarefaction wave moving to the left and a highly relativistic shock wave moving to the right with velocity 
$ \sim 0.94$ (in $c=1$ units). The solution domain is $\left(x,t \right) = \left[0,1.5 \right] \times \left[0,0.8 \right]$.

\begin{figure*}
    \centering
    \includegraphics[width=0.93\textwidth]{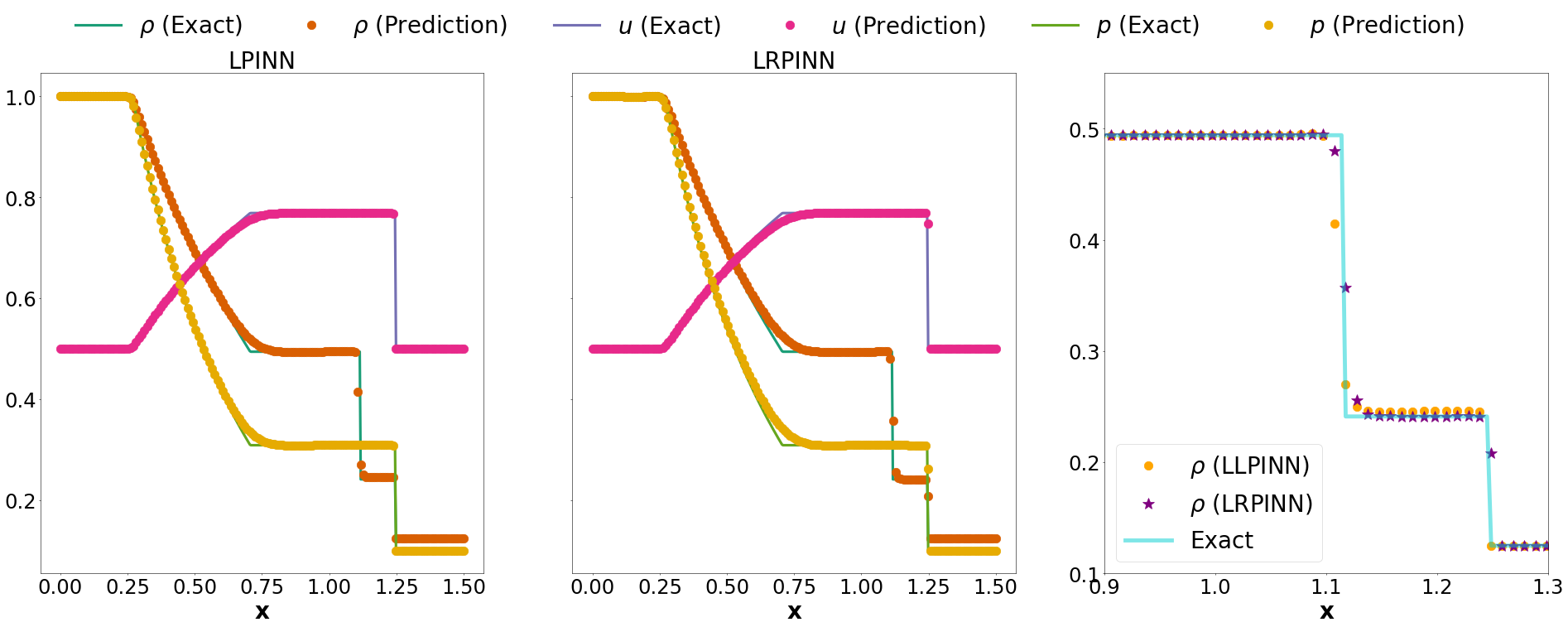}
    \caption{Same as Figure \ref{fig:lax_euler_predictions} for the relativistic Sod shock tube problem.}\label{fig:predictions_problem_4}
\end{figure*}

Figure \ref{fig:predictions_problem_4} shows the LLPINN and LRPINN predictions. A slight discrepancy in the density appears again for the LLPINN between the contact and the shock waves, as noted in the right panel of the corresponding right panel.

\begin{figure*}
    \centering
    \includegraphics[width=0.93\textwidth]{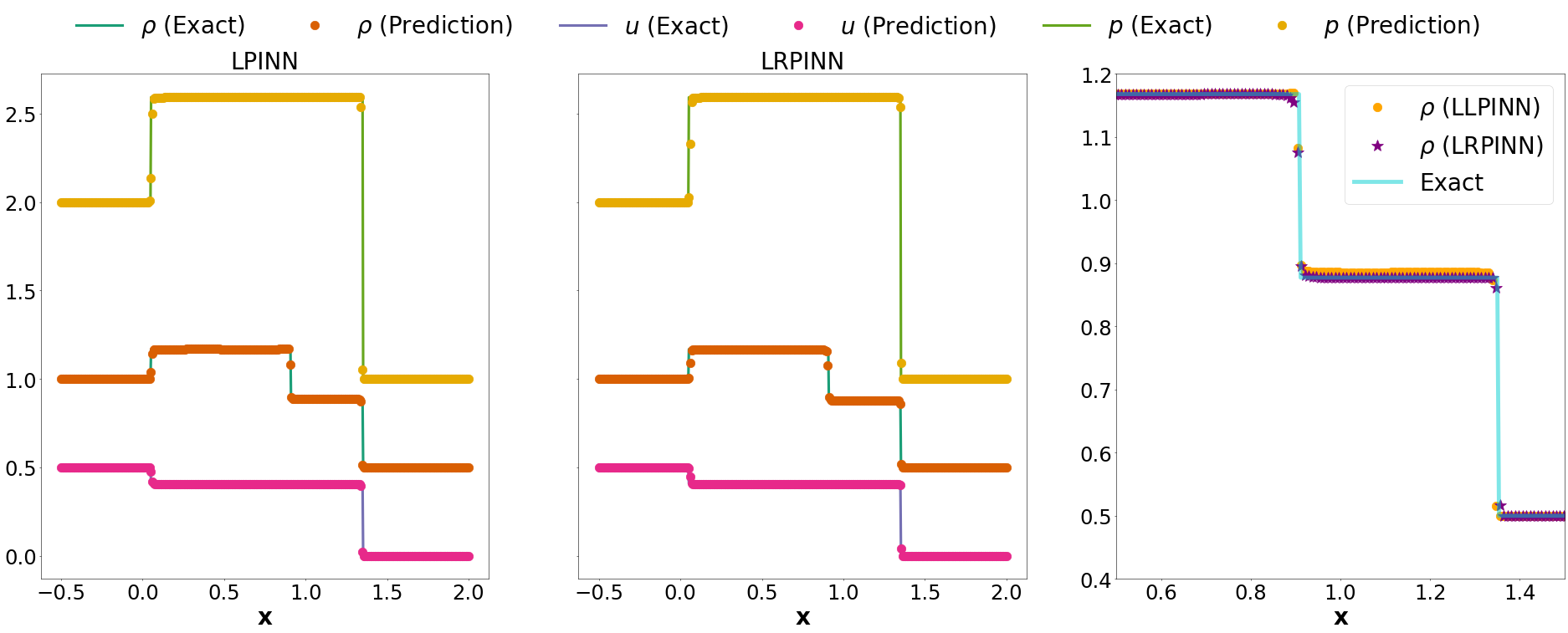}
    \caption{Same as Figure \ref{fig:lax_euler_predictions} for the relativistic double shock problem.}\label{fig:predictions_problem_5}
\end{figure*}

\paragraph{\textbf{Problem 5: Relativistic double shock wave}}
To end with relativistic hydrodynamics, consider the initial conditions
\begin{equation*}
    (\rho,u,p)=
    \begin{cases}
        &(1.0,0.5,2.0), \quad x < 0.5, \\
        &(0.5,0.0,1.0), \quad x>0.5,
    \end{cases}
\end{equation*}
which are similar to the ones presented in the \emph{Problem 3} of \cite{ferrer2024gradient}, but with a higher pressure value at the left state, which provides the solution with a stronger \emph{right-moving} and specific enthalpy values between $h=6$ and $h \sim 8.4$.  Here, the adiabatic index $\gamma = 5/3$, and the domain is $\left(x,t\right) \in \left[-0.5,2.0 \right] \times \left[0,1 \right]$. 

Figure \ref{fig:predictions_problem_5} shows the predictions of LLPINNs and LRPINNs for this case. However, in the right panel of Figure \ref{fig:predictions_problem_5}, we observe that the LLPINN incorrectly predicts the density value between the contact wave and the right-moving shock wave, while the LRPINN reproduces the correct result.

\subsection{2D problems}

For the mutlidimensional cases, the LLPINN and LRPINN approaches are summarized in Algorithms \ref{alg:LLPINNs2D} and \ref{alg:lrpinns2D}, respectively. We also use a gradient-annihilated factor defined as follows
\begin{equation}
    \alpha = \frac{1}{1 + a\sqrt{\left(\frac{\partial u}{\partial x} \right)^2 + \left(\frac{\partial v}{\partial y} \right)^2}}.
\end{equation}
Table \ref{tab:hyperparameters} shows the number of layers and neurons of the neural network and the L(R)PINN parameters employed for each problem.
The loss function is again computed using expression \eqref{eq:loss}, with weights set to $\omega_R = 1$, and $\omega_I = 10$. 

The training set now consists of random points $(x,y,t) \in \Omega_R \times \Omega_I$ following a uniform distribution, and again are randomly changed every $500$ iterations at most. 

\begin{figure*}
    \centering
    \includegraphics[width=0.85\textwidth]{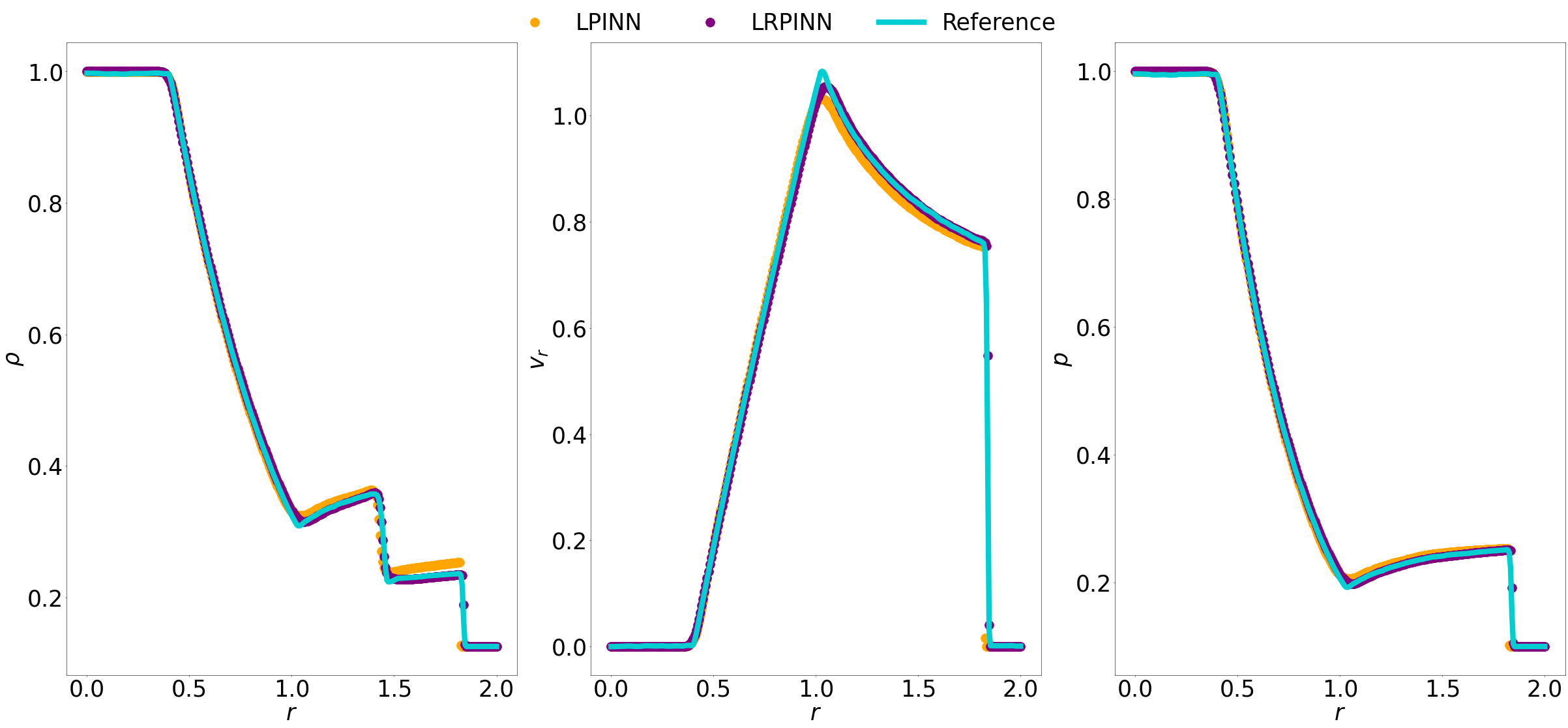}
    \caption{Predictions of the density, velocity and pressure profiles of the cylindrical Sod shock tube problem. Here, $v_r = \sqrt{u^2 + v^2}$ denotes the velocity in the radial direction.}
    \label{fig:2d_cylindrical}
\end{figure*}

\paragraph{\textbf{Problem 6: Cylindrical Sod shock tube}} 
We consider with the analog of the Sod tube in cylindrical symmetry\cite{OMANG2006391,BERNARDCHAMPMARTIN2013170}
\begin{equation*}
    (\rho,u,v,p)=
    \begin{cases}
        &(1.0,0,0,1.0), \quad x ^2 + y^2< 1, \\
        &(0.125,0,0,0.1), \quad x^2 + y^2 >1.
    \end{cases}
\end{equation*}

with solution domain $(x,y,t) \in \left[0,2 \right]^2 \times \left[0,0.5\right]$. The structure of the solution is similar to the 1D Sod shock tube in cartesian coordinates that we have seen in Problem 2, as we have a rarefaction wave, followed by a contact wave, and finally a shock wave. However, due to the geometrical source term that appears in the cylindrical version of the 1D system of conservation laws, noticeable differences appear in the solution profiles (see Figure  \ref{fig:2d_cylindrical}).

\begin{figure*}
    \centering
    \includegraphics[width=0.85\textwidth]{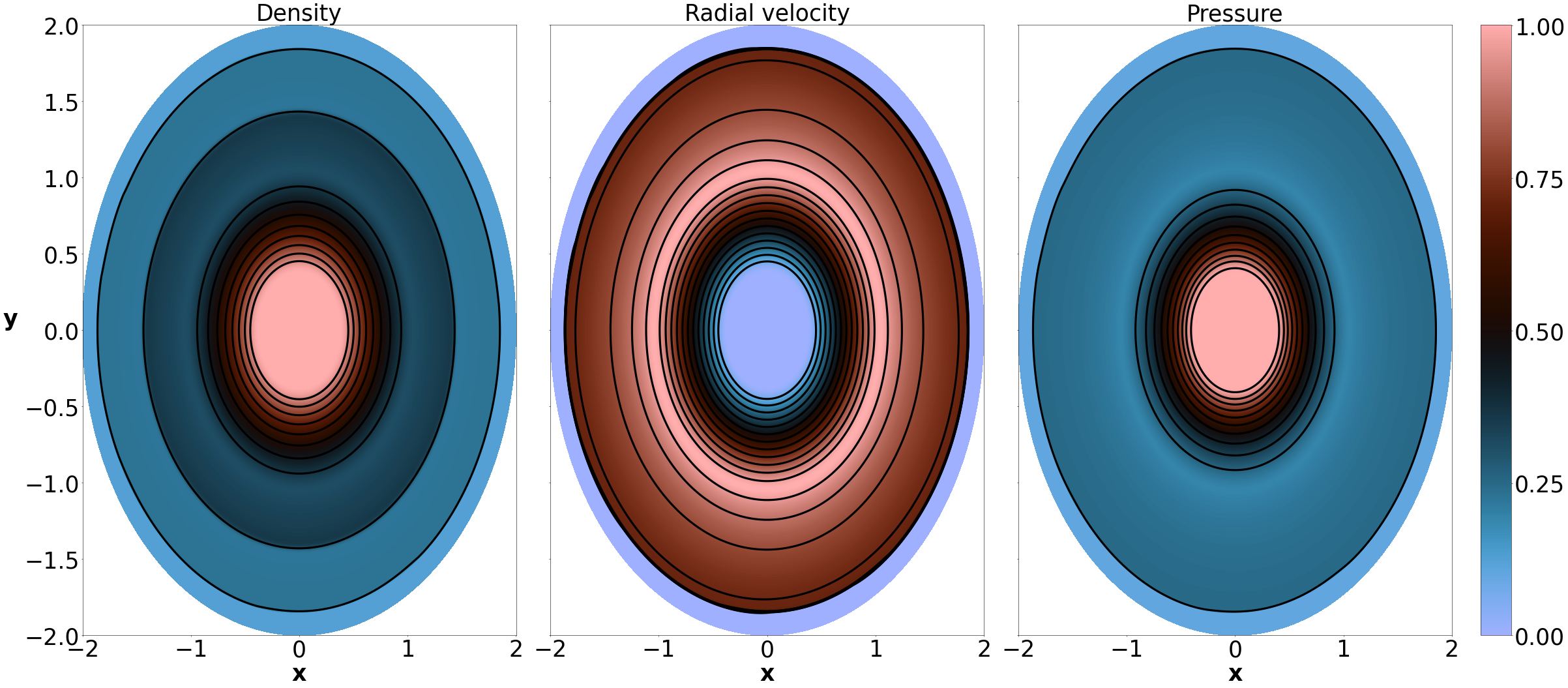}
    \caption{Colormaps with overimposed contours of the LRPINN predictions of the density, the radial velocity and the pressure. The radial velocity has been normalized to its maximum value to have the same color scale in the three panels.}\label{fig:2d_cylindrical_colormap}
\end{figure*}

Figure \ref{fig:2d_cylindrical} shows the predicted profiles along one radial direction, together with a reference numerical solution (since no analytical solution is available for this case\footnote{The solution closely resembles that presented in \cite{OMANG2006391} (see their Figure 11).}), obtained by using a 1D HLLC approximate Riemann solver\cite{hydrofv} written in cylindrical coordinates with $300$ radial points. Both LLPINN and LRPINN accurately capture the shock location. As expected, LRPINN provides a more precise representation of the variable jumps, particularly in the density. Additionally, the velocity prediction from LLPINN appears significantly more diffused in the region near its peak. 
Figure \ref{fig:2d_cylindrical_colormap} shows colormaps of the LRPINN prediction, with different contour levels overimposed, to illustrate that the neural network correctly maintains the cylindrical symmetry of the problem.

Finally, we perform the same error analysis as that presented in the 1D problem section (see Table~\ref{tab:errors}) for this case. Since there is no analytical solution available, we compare our results with a reference solution obtained with a 1D HLLC\cite{hllc} scheme in cylindrical coordinates, with a resolution of 1600 points. Table~\ref{tab:errors2d} reports the relative $L^2$ error norms for the density, the radial velocity $v_r = \sqrt{u^2 + v^2}$, and the pressure computed using the 1D cylindrical HLLC approximate Riemann solver at various resolutions, alongside the corresponding quantities obtained with the LRPINN. As we can see, taking as reference the 1D HLLC solution with 1600 points, the 2D LRPINN solution achieves a level of precision comparable to that of the HLLC scheme with a resolution of 400 points.

\begin{table}[h!]
    \centering
    \begin{tabular}{ccccc}
         Method & Points & $L^2_\rho$ & $L^2_{v_r}$ & $L^2_p$ \\ \hline 
         1D HLLC  & $50$ & $0.027$ & $0.074$ & $0.028$ \\
         1D HLLC  & $100$ & $0.015$ & $0.062$ & $0.013$ \\
         1D HLLC  & $200$ & $0.014$ & $0.043$ & $0.009$        \\
         1D HLLC  & $400$ & $0.006$ & $0.027$ & $0.006$       \\
         LRPINN & - &  $0.009$ & $0.014$ & $0.007$
    \end{tabular}
    \caption{Relative $L^2$ error norms of the differences with the numerical reference solution for the density, radial velocity, and pressure predictions obtained using a 1D HLLC\cite{hydrofv} scheme in cylindrical coordinates at different spatial resolutions, and with the LRPINN.}
    \label{tab:errors2d}
\end{table}

\paragraph{\textbf{Problem 7: 2D Riemann problem}} 

Next, we consider the 2D version of the Riemann problem. We start in particular with the following initial condition
\begin{equation*}
    \mathbf{w}=
    \begin{cases}
        &(1.1,0,0,1.1), \quad \left(x,y \right) \in \left[0,0.5 \right] \times \left[0,0.5 \right], \\
        &(0.51,0.89,0.89,0.35), \quad \left(x,y \right) \in \left[-0.5,0 \right] \times \left[0,0.5 \right], \\
         &(1.1,0.89,0.89,1.1), \quad \left(x,y \right) \in \left[-0.5,0 \right] \times \left[-0.5,0 \right], \\
         &(0.51,0,0.89,0.35), \quad \left(x,y \right) \in \left[0.5,0 \right] \times \left[-0.5,0 \right]
    \end{cases}
\end{equation*}
where $\mathbf{w} \equiv \left(\rho,u,v,p\right)$, which corresponds to the initial configuration of Problem 4 of  \cite{Kurganov_Tadmor}. The domain of the solution is now $(x,y,t) \in \left[-0.5,0.5\right]^2 \times [0,0.25]$.
The solution to this problem consists of four different and interacting shock waves, corresponding to the four interfaces of the initial condition moving in different directions. 

\begin{figure*}
    \centering
    \includegraphics[width=0.93\textwidth]{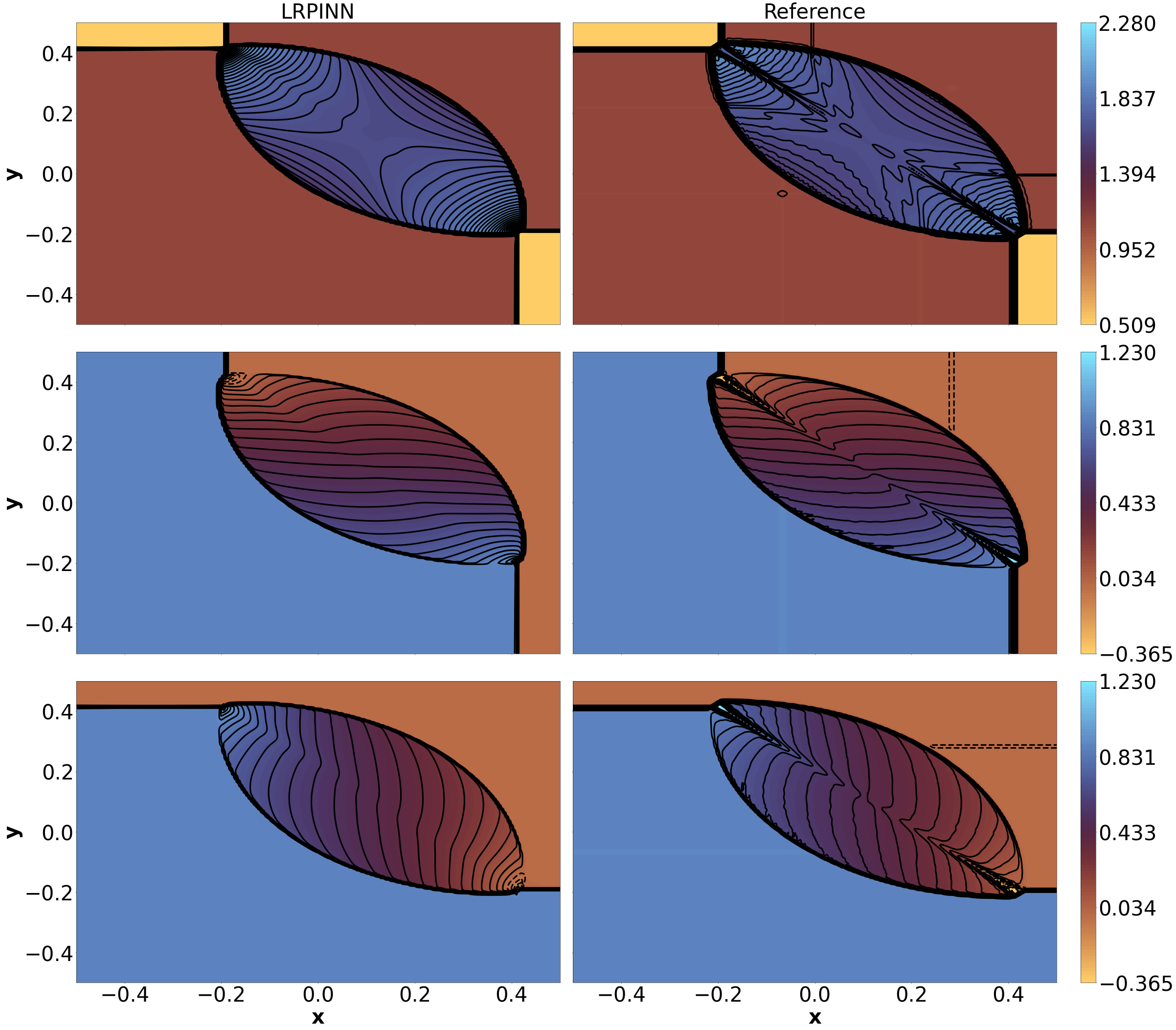}
    \caption{Colormaps with overimposed contours of the LRPINN (left) and a numerical 2D Riemann solver (right) of the density (top panels), horizontal (middle panels), and vertical (bottom panels) velocity components.}\label{fig:2d_roe_p4}
\end{figure*}

The left panels of Figure \ref{fig:2d_roe_p4} show the density and the velocity predictions obtained using the LRPINN, while the right panels show the reference solution computed with a numerical 2D Riemann solver \cite{hypar2023}, based on a fifth-order WENO scheme combined with a third-order strong stability preserving Runge-Kutta method. We observe that LRPINN accurately captures the formation of the various shock waves and exhibits sharper transitions compared to the numerical solution, as indicated by the denser contour levels in these regions. However, it significantly smooths the small-scale structures in the central part of the domain. Addressing multi-scale phenomena with PINNs remains particularly challenging, as these models tend to smooth out high-frequency components of the solution, which in turn negatively affects the overall convergence of the neural network \cite{wang2021eigenvector,SAPINNs,moseley2023fbpinns,Dolean2024}. 

\paragraph{\textbf{Problem 8: 2D Riemann problem}}

We end with a solution that contains the three types of waves (that is, rarefaction, contact, and shock waves), which corresponds to the following initial condition
\begin{equation*}
    \mathbf{w}=
    \begin{cases}
        &(0.53,0.1,0.1,0.4), \quad \left(x,y \right) \in \left[0,0.5 \right] \times \left[0,0.5 \right], \\
        &(1.02,-0.62,0.1,1), \quad \left(x,y \right) \in \left[-0.5,0 \right] \times \left[0,0.5 \right], \\
         &(0.8,0.1,0.1,1), \quad \left(x,y \right) \in \left[-0.5,0 \right] \times \left[-0.5,0 \right], \\
         &(1,0.1,0.83,1), \quad \left(x,y \right) \in \left[0.5,0 \right] \times \left[-0.5,0 \right]
    \end{cases}
\end{equation*}
where, again, $\mathbf{w} \equiv \left(\rho,u,v,p\right)$. This problem corresponds to Problem 16 of \cite{Kurganov_Tadmor}. The solution domain is the same as before, that is, $(x,y,t) \in \left[-0.5,0.5\right]^2 \times [0,0.25]$. The solution features a shock wave between the first and fourth quadrants, a rarefaction wave between the first and second quadrants, and two contact discontinuities between the second and third quadrants and the third and fourth quadrants. 

\begin{figure*}
    \centering
    \includegraphics[width=0.85\textwidth]{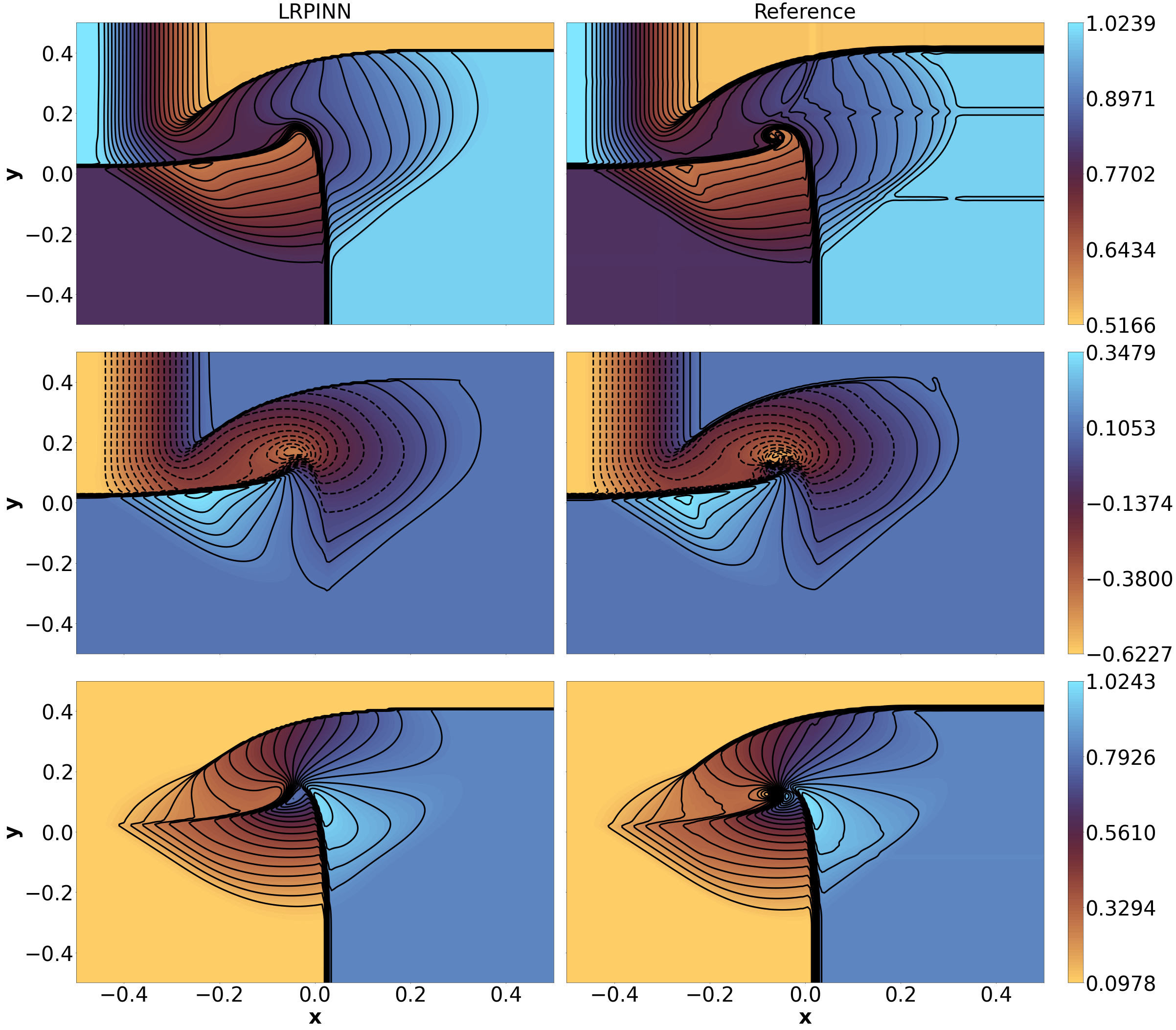}
    \caption{Same as Figure \ref{fig:2d_roe_p4} for Problem 8.}\label{fig:2d_roe_p16}
\end{figure*}

\begin{figure*}
    \centering
    \includegraphics[width=0.85\textwidth]{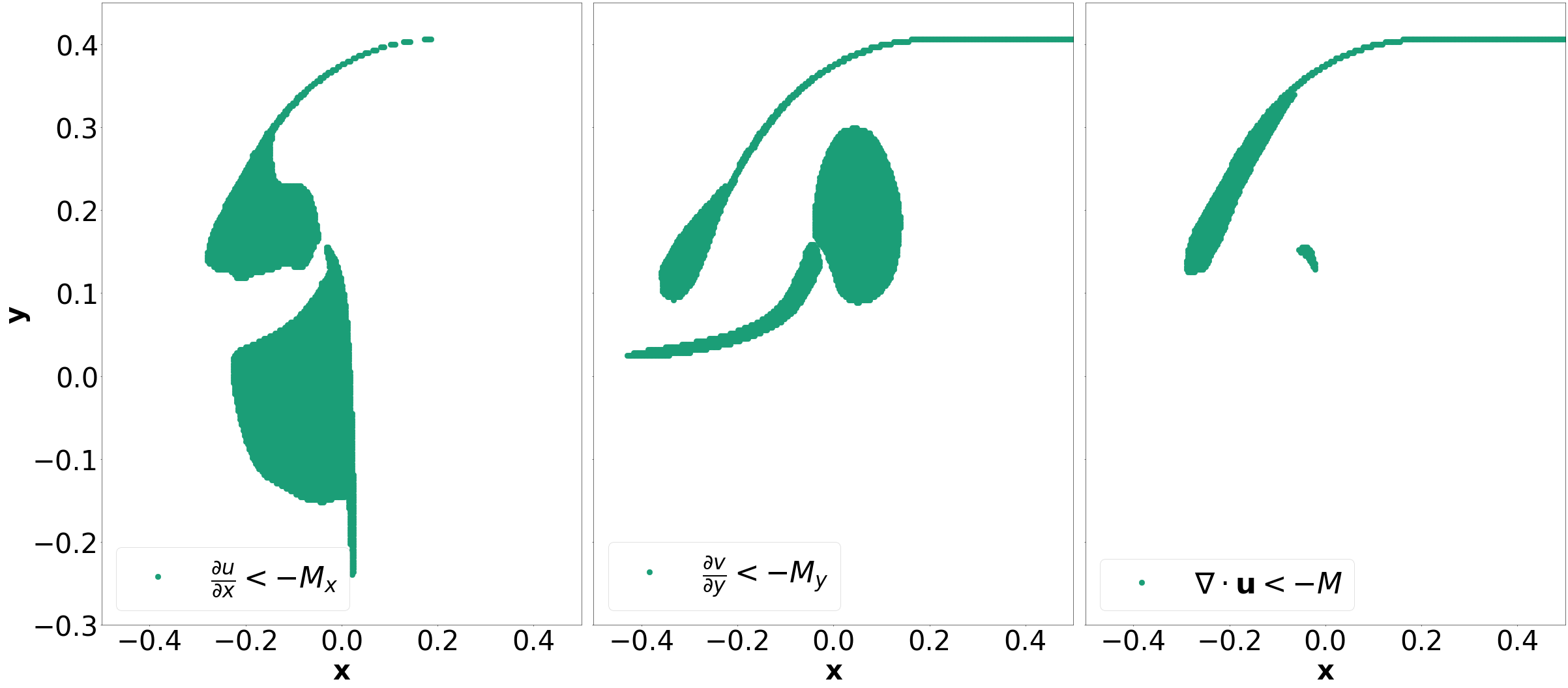}
    \caption{Points obtained from a test set (a $300 \times 300$ grid) that satisfy: $\frac{\partial u}{\partial x} < -M_x$ (left panel), $\frac{\partial v}{\partial y} < -M_y$ (middle panel) and $\nabla \cdot \mathbf{u} < -M$, where the parameters $M_x$, $M_y$ and $M$ are the ones specified at Table \ref{tab:hyperparameters} for this problem.}\label{fig:2d_shock_points}
\end{figure*}

The left panels of Figure \ref{fig:2d_roe_p16} show the density and the velocity LRPINN predictions, whereas the right panels show the corresponding quantities predicted by the same numerical solver used at Problem 7.  Once again, we observe that the discontinuities associated with the contact waves, and especially the shock wave, seem more sharply resolved by the LRPINN compared to the numerical solution. However, the vortex structure in the central region of the domain is noticeably smoothed out. 

To conclude this section, we explain why the divergence, rather than individual gradients in each coordinate direction, is essential for identifying shock formation regions.
To illustrate this, we evaluate the neural network solution at the final time
and select the points that meet the specified criteria:
\begin{itemize}
    \item $\frac{\partial u}{\partial x} < -M_x$
    \item $\frac{\partial v}{\partial y} < -M_y$
    \item $\nabla \cdot \mathbf{u} < -M$
\end{itemize}
where the values $M_x$, $M_y$ and $M$ are listed in Table \ref{tab:hyperparameters}). Figure \ref{fig:2d_shock_points} shows the regions that satisfy each of the three inequalities. 
The divergence condition accurately pinpoints the region where a true shock forms
(the jump in the $y$ component of the velocity in the top-right region) , whereas the other two conditions, if considered separately, could misguide the algorithm to apply the modified Jacobian in regions with large gradients, but not an actual shock wave.

\section{Conclusions and future work}

In this paper, we sought to enhance and broaden the capability of
PINNs to provide solutions for discontinuous problems in hydrodynamics and relativistic hydrodynamics. 
The classical PINN method, which trains the neural network on the residuals of the conservation laws in their original form, fails to produce accurate results in most cases examined here. We found that this problem persists even when the network is pre-initialized with a reasonable viscous approximation.

In this work, we have proposed a more general extension of LLPINNs (Algorithm \ref{alg:LLPINNs}) that does not need any prior information about the shock waves present in the solution. Instead, we dynamically predict the shock speeds using information of neighbor points, ensuring that the Rankine–Hugoniot relations \eqref{eq:RH} and the entropy conditions \eqref{eq:lax_entropy} are satisfied by construction.

In addition, a detailed analysis of LLPINNs revealed that, in some cases, even when the shock velocity was accurately predicted, the jump in the conserved variables was not correctly captured. We attribute this issue to the fact the modification applied to the Jacobian matrix in the original method does not guarantee conservation across shocks, ultimately leading to a violation of global conservation laws that grows over time (see Figure \ref{fig:mass_residuals_lax}). To address this limitation, we propose a modification in which the Jacobian matrix is replaced, in potential shock regions, by the Roe matrix \cite{Roe1997}. 
Our new method, termed LRPINNs (see Algorithm \ref{alg:lrpinns}), robustly ensures that both shock velocities and jump conditions are met. The strategy used in this work utilizes the Roe-approximate solution of the Riemann problem, but it can be readily adapted or extended to other approximate solvers (or the exact solution when available).
For example, it would suffice to diagonalize the Jacobian of the system, replace the eigenvalues with those consistent with the chosen approximate Riemann solver, and then reconstruct the modified Jacobian in the original basis with the eigenvector matrix (see e.g. \cite{Aloy1999} for similar procedures in classical methods).

We have also extended the scope of LLPINN and LRPINN algorithms to tackle two-dimensional Riemann problems (see Algorithms \ref{alg:LLPINNs2D} and \ref{alg:lrpinns2D}). 
The identification of potential shock regions is generalized from the one-dimensional case by evaluating the divergence of the velocity field. 
The Jacobian matrices in each coordinate direction are then modified using a \emph{dimensional splitting} strategy, where each one-dimensional corrections are applied independently along each spatial direction. 
We tested our method on complex 2D problems featuring shocks and other discontinuities, yielding results consistent with those from one-dimensional cases. 
The results were benchmarked against a high-order WENO solver. 
LRPINNs effectively captured shock waves, producing sharper transitions than the WENO method, although some regions with high vorticity appeared noticeably smoothed. 
A critical focus for future work is to devise strategies that improve the resolution of small-scale structures, crucial for turbulence, while maintaining LRPINNs' good performance in capturing shock formations.

\begin{acknowledgments}
JFU is supported by the predoctoral fellowship ACIF 2023, cofunded by Generalitat Valenciana and the European Union through the European Social Fund.
We acknowledge support from the Prometeo excellence programme grant
CIPROM/2022/13 funded by the Generalitat Valenciana. We also acknowledge 
support through the grant PID2021-127495NB-I00 funded by MCIN/AEI/10.13039/501100011033 and by the European Union, as well as the Astrophysics and High Energy Physics programme of the Generalitat Valenciana ASFAE/2022/026 funded by MCIN and the European Union NextGenerationEU (PRTR-C17.I1).

\end{acknowledgments}

\nocite{*}
\bibliography{aipsamp}

\end{document}